\documentclass[useAMS,usenatbib]{mn2e}
\usepackage{amssymb,epsfig,natbib,rotating,captcont,capt-of,longtable}

\title[Sher\,25 and its intriguing hourglass nebula]
{The blue supergiant Sher\,25 and its intriguing hourglass nebula}
\author[M. A. Hendry {\rm et al.}]{M.~A. Hendry$^{1,2}$, S.~J. Smartt$^1$, E.~D. Skillman$^3$, C.~J. Evans$^4$, C. Trundle$^1$, \newauthor
 D. J. Lennon$^{5,6}$,  P. A. Crowther$^7$, I. Hunter$^1$, \\
$^1$Astrophysics Research Centre, School of Maths and Physics,Queen's University Belfast, Belfast BT7 1NN, UK\\
$^2$Institute of Astronomy, University of Cambridge, Madingley Road,
        Cambridge CB3 0HA\\
$^3$Astronomy Department, University of Minnesota, 116 Church Street SE, Minneapolis, MN 55455, USA\\
$^4$UK Astronomy Technology Centre, Royal Observatory Edinburgh, Blackford Hill, Edinburgh EH9 3HJ\\
$^5$Space Telescope Science Institute, 3700 San Martin Drive, Baltimore, MD 21218, USA\\
$^6$Instituto de Astrof{\'i}sica de Canarias, 38200 La Laguna, Tenerife, Spain\\
$^7$Department of Physics \& Astronomy, University of Sheffield, Hicks Building, Hounsfield Rd, Sheffield, S3 7RH, UK}

\date{Accepted: 2008 March 18}

\pagerange{\pageref{firstpage}--\pageref{lastpage}} \pubyear{2007}

\def\LaTeX{L\kern-.36em\raise.3ex\hbox{a}\kern-.15em
  T\kern-.1667em\lower.7ex\hbox{E}\kern-.125emX}
\def \aj {AJ}
\def \mnras {MNRAS}
\def \apj {ApJ}
\def \apjl {ApJL}
\def \aap {A\&A}
\def \nat {Nature}

\def \pasp {PASP}

\def \jrasc {JRASC}

\def \aaps {A\&AS}

\newcommand{\bv}{\mbox{$B\!-\!V$}}
\newcommand{\degree}{\mbox{$^\circ$}}
\newcommand{\msun}{\mbox{M$_{\odot}$}}
\newcommand{\rsun}{\mbox{R$_{\odot}$}}
\newcommand{\kms}{\mbox{$\rm{km}\,s^{-1}$}}

\DeclareMathAlphabet{\mathsc}{OT1}{cmr}{m}{sc}
\def\testbx{bx}%
\DeclareRobustCommand{\ion}[2]{%
\relax\ifmmode
\ifx\testbx\f@series
{\mathbf{#1\,\mathsc{#2}}}\else
{\mathrm{#1\,\mathsc{#2}}}\fi
\else\textup{#1\,{\mdseries\textsc{#2}}}%
\fi}
\newcommand{\ha} {\mbox{H$\alpha$}}
\newcommand{\hb} {\mbox{H$\beta$}}
\newcommand{\hg} {\mbox{H$\gamma$}}
\newcommand{\hd} {\mbox{H$\delta$}}

\newcommand{\Hii} {\ion{H}{ii}}
\newcommand{\Hei} {\ion{He}{i}}
\newcommand{\Heii} {\ion{He}{ii}}

\newcommand{\Nii} {[\ion{N}{ii}]}

\newcommand{\Oii} {[\ion{O}{ii}]}
\newcommand{\Oiii} {[\ion{O}{iii}]}
\newcommand{\Sii} {[\ion{S}{ii}]}
\newcommand{\Siii} {[\ion{S}{iii}]}
\newcommand{\SiIII} {\ion{Si}{iii}}
\newcommand{\Neiii} {[\ion{Ne}{iii}]}
\newcommand{\Ariii} {[\ion{Ar}{iii}]}
\newcommand{\Ariv} {[\ion{Ar}{iv}]}
\newcommand{\Cliii} {[\ion{Cl}{iii}]}

\begin{document}

\label{firstpage}

\maketitle

\begin{abstract}

The blue supergiant Sher\,25 is surrounded by an asymmetric,
hourglass-shaped circumstellar nebula. Its structure and dynamics have
been studied previously through high-resolution imaging and
spectroscopy, and it appears dynamically similar to the ring structure
around SN\,1987A. Here we present long-slit spectroscopy of the
circumstellar nebula around Sher\,25, and of the background nebula of
the host cluster NGC\,3603. We perform a detailed nebular abundance
analysis to measure the gas-phase abundances of oxygen, nitrogen,
sulphur, neon and argon.  The oxygen abundance in the circumstellar
nebula ($12 + \log {\rm O/H} = 8.61\pm0.13$\,dex) is similar to that
in the background nebula ($8.56\pm0.07$), suggesting the composition
of the host cluster is around solar.  However, we confirm that the
circumstellar nebula is very rich in nitrogen, with an abundance of
$8.91\pm0.15$, compared to the background value of $7.47\pm0.18$. A new
analysis of the stellar spectrum with the {\sc fastwind} model
atmosphere code suggests that the photospheric nitrogen and oxygen abundances in
Sher\,25 are consistent with the nebular results. While the nitrogen
abundances are high, when compared to stellar evolutionary models they
do not unambiguously confirm that the star has undergone convective
dredge-up during a previous red supergiant phase.  We suggest that the
more likely scenario is that the nebula was ejected from the star
while it was in the blue supergiant phase. The star's initial mass was
around 50\msun, which is rather too high for it to have had a
convective envelope stage as a red supergiant. Rotating stellar models
that lead to mixing of core-processed material to the stellar surface
during core H-burning can quantitatively match the stellar results
with the nebula abundances.
\end{abstract}

\section{Introduction}

Sher\,25 \citep{1965MNRAS.129..237S} is an evolved blue supergiant
(BSG), with a $V$-band magnitude of 12.28 and \bv\ colour of 1.38
\citep{1978A&A....63..275V}. It has a spectral type of B1.5~Ia
\citep{1983A&A...124..273M}, similar to Sk~\mbox{$-69\degree202$}, the
progenitor of SN\,1987A, which was a B3~I supergiant when it exploded
\citep{1989A&A...219..229W,1987A&A...177L..25P,1987Natur.328..318G}. Sher\,25
is situated about 20$''$ north of the trapezium-like system
HD\,97950, at the core of the young cluster in the centre of the giant
\Hii\ region NGC\,3603. Figure~\ref{fig:EW} (lower panel) shows part of a {\it
HST} WFPC2 image\footnote{Image credit: Wolfgang Brandner (JPL/IPAC),
Eva K. Grebel (U. Wash.), You-Hua Chu (UIUC), NASA} of this giant
Galactic emission nebula with Sher\,25 in the top left-hand corner,
clearly showing its complicated emission-line nebula. Sher\,25 was
first reported by \citet{1965MNRAS.129..237S}, who used photometric
observations to ascertain the distance to NGC\,3603 and its reddening.
\citet{1997ApJ...475L..45B} discovered the ring-shaped 
nebula and apparent bipolar outflows serendipitously
during a high-resolution spectral study of emission-line knots in the
giant \Hii\ region. The authors likened the circumstellar nebula to
that of SN\,1987A, where the rings existed before the supernova
explosion and hence appear to have been ejected by the progenitor
star.

\citet{1997ApJ...475L..45B}, using high-resolution echelle spectra,
showed that the ring and the outflow to the north-east had a high
\Nii/H$\alpha$ ratio, compared to the background \Hii\ region. In an
\Oiii/H$\beta$ versus \Nii/H$\alpha$ diagram, the north-east outflow
filament was clearly situated outside the location of \Hii\ regions
and supernova remnants. The authors interpreted this as an enhanced N
abundance and concluded that the bipolar filaments consisted of
stellar material that had been enriched by the CNO cycle.  On this
basis they further suggested that Sher\,25 was an evolved BSG that had
passed through the red supergiant (RSG) phase, possibly a twin of the
progenitor of SN\,1987A (Sk~\mbox{$-69\degree202$}).  Note that
enhanced N in a stellar photosphere or circumstellar nebulae does not
necessarily imply post-RSG status
\cite[see, e.g.,][]{2001ApJ...551..764L,2000A&A...361..101M}.
The comparison with Sk~\mbox{$-69\degree202$} is attractive, but one
should note that the two central stars are rather different in
luminosity and mass, Sher\,25 (40-60\msun) is more massive than 
Sk~\mbox{$-69\degree202$} (15-20\msun). 

In a second paper,
\citet{1997ApJ...489L.153B} presented a study of the ring and bipolar
outflows to investigate their physical structure and dynamics. Using
long-slit spectroscopic mapping of Sher\,25's nebula, they uncovered an
asymmetric hourglass shape, with the inner ring at the waist of the
hourglass. Their new observations confirmed the inclination angle from
\citet{1997ApJ...475L..45B}, but found different expansion
velocities. \citet{1997ApJ...475L..45B} found far larger velocities in 
the bipolar outflows than in the ring, indicating
that they were ejected at different times. However,
\citet{1997ApJ...489L.153B} found more comparable velocities and
derived dynamical ages of 6,560 and 6,700~years for the ring and bipolar
lobes. Because of the comparable ages of the ring and the lobes, the
authors concluded that {\em both} the ring and the bipolar lobes were
formed from the same brief and violent mass ejection $\sim$6,630 years
ago. \citet{1997ApJ...475L..45B,1997ApJ...489L.153B} also carried out
a detailed comparison of Sher\,25's hourglass nebula with the ring
structure around SN\,1987A.  \citet{1997ApJ...475L..45B} suggested that
the main differences could be attributed to the
different metallicities of the regions and the main-sequence
masses of the stars. They also suggested that the bipolar nebulae
represented the first members of a new class of nebula around BSGs in
the final stages of their evolution, somewhere between luminous blue
variables (LBVs) and planetary nebulae.  \cite{2007AJ....133.1034S} has
recently discovered another Galactic BSG (HD\,168625) with a
similar circumstellar ring structure.  Smith suggests that this could
have come from an LBV-type eruption rather than being formed through a
RSG-BSG wind interaction, and by implication that perhaps the SN\,1987A
rings were formed in a similar manner. Subsequently,
\cite{2007AJ....134..846S} announced two further discoveries of
ring nebulae in the Carina Nebula, one of which encircles an early
B-type supergiant. They speculate that these nebulae may be much more
common than we have previously assumed.

\citet{1989ApJ...336..429F} suggested that Sk~\mbox{$-69\degree202$} 
had passed through a RSG phase because of large over-abundances of
nitrogen (compared to carbon and oxygen) in early UV spectra, although
their uncertainties were large.  The nebular spectroscopy from
\citet{1997ApJ...475L..45B,1997ApJ...489L.153B} seems to suggest that
Sher\,25 has also undergone a RSG phase, and is currently at a similar
evolutionary stage to that of Sk~\mbox{$-69\degree202$} before it exploded.
In this scenario, Sher\,25 evolves from a main-sequence O-type star to
a RSG, at which point it undergoes convective `dredge-up' of
CNO-processed material, thereby dramatically changing the surface
abundances -- a major mass-ejection event at this point could then
explain the N-enriched nebula.  Sher\,25 would then evolve back across
the H--R diagram to greater temperatures via a `blue loop', ultimately
returning to the red or continuing to evolve to hotter temperatures as
a Wolf-Rayet star.  An alternative scenario to account for the
N-enrichment of Sher\,25 is that the CNO-processed material was mixed
from the core into the envelope via rotationally-induced mixing, while
the star was on (or near) the main sequence.  Given the observed
range of rotational velocities, the CNO abundances seen in Galactic B-type supergiants
\citep{1999A&A...349..553M,2006A&A...446..279C} are consistent with
the predictions of rotationally-induced mixing; the results for Sher\,25 were 
suggested to be similar to this population by
\cite{arXiv:astro-ph/0606717}.  We also note that the massive-star population 
in NGC\,3603 appears to be predominantly coeval (with an age of
1-2\,Myr), but that Sher\,25, and one O-type supergiant, are likely to
be slightly older \citep[$\sim$4\,Myrs;][]{arXiv:0712.2621}.

\citet[][hereafter Paper~I]{2002A&A...391..979S} presented high-resolution optical 
spectroscopy of Sher\,25 alongside a model photosphere and unified
stellar wind analysis. They determined atmospheric parameters,
the mass-loss rate and photospheric abundances for C, N, O, Mg and Si and
compared them with other Galactic B-type stars. They found that
Sher\,25 was not extreme or abnormal in terms of its photospheric N
abundance. They also compared the C/N and N/O abundance ratios to
surface abundances predicted by stellar evolutionary calculations,
which assumed that the star had been through a RSG phase with
convective dredge-up. The N/O abundance ratio was found to be
inconsistent with the star having been a RSG and therefore the nebula
was likely to have been expelled during the BSG phase
instead. However, the results were found to be consistent with some
degree of rotational mixing while the star was near the main
sequence. The wind analysis also suggested that Sher\,25 has a
relatively normal mass-loss rate in comparison with other Galactic
B-type supergiants. Paper~I also inspected the
spectra for signs of binarity, but the exposures had insufficient
temporal coverage over which to search for significant variations.

At present, the data from the star and the nebula of Sher\,25 appear to
be in conflict. Paper~I did not find any evidence
to suggest that the nebula was ejected during a previous RSG phase, or
that Sher\,25 had in fact been a RSG at all.  Moreover, no evidence for
blue loops has been found in studies of B-type supergiants
\citep{1999A&A...349..553M}, nor in LBVs \citep{2001ApJ...551..764L}.
However, although the higher \Nii/H$\alpha$ ratios in comparison to the
background ratios found by \citet{1997ApJ...475L..45B} are indicative
of N enrichment from the CNO cycle, it is not conclusive. No
quantitative abundance analysis was carried out by
\citet{1997ApJ...475L..45B,1997ApJ...489L.153B}, as the spectra
did not have sufficient wavelength coverage to determine electron
densities and temperatures. In order to ascertain whether the
hourglass nebula of Sher\,25 is consistent with it having undergone a
blue loop, a detailed abundance analysis is required.

The evolutionary status of Sher\,25 has important implications for the
progenitor of SN\,1987A. The current theory of the evolution of single
stars cannot account for several features of Sk~\mbox{$-69\degree202$}
and SN\,1987A, i.e., the fact that Sk~\mbox{$-69\degree202$} exploded as a
BSG, the triple ring nebula, nor the chemical anomalies observed,
principally the enhancement of He and CNO products
\citep{2003fthp.conf...13P}. This raises the question, did
Sk~\mbox{$-69\degree202$} actually go through a RSG phase? Was it a
completely unique star?  The observations have been reproduced
by recent models of binary evolution that invoke the `slow merger' of a
RSG with a mass of 15--20\,\msun\ with a companion of 1--2\,\msun,
depending on the evolutionary state of the primary at the onset of the
merger \citep{2003fthp.conf...19I,2003fthp.conf...13P}. \citet{2005ASPC..342..194M}
then demonstrated how this can lead to the observed triple ring
structure. Could the nebula around Sher\,25 also be attributed to the
presence of a companion star?  A detailed abundance analysis could
also indicate the origin of the nebula. We would assume that the
abundances in the nebula would be the same as those in the photosphere
of the star if it had been ejected in the near past. However, if the
abundances showed real differences then it would indicate that the
nebula was not the result of a mass ejection from Sher\,25, but perhaps
the result of the destruction of (or a severe mass-loss event from) an unseen
companion.

In this paper we attempt to answer some of the questions that have
been raised by the previous analyses. New nebular and stellar
observations are presented in Section~\ref{sec:obs}, followed by
abundance analyses in Section~\ref{sec:neb} and a discussion of a
potential binary companion in Section~\ref{sec:bin}. The implications
of both the nebular analysis and the search for a companion are
summarised in Section~\ref{sec:conc}.

\section{Observations}\label{sec:obs}

\subsection{Nebular observations}
\label{sec:longslitobs}
\begin{figure*}
  \begin{minipage}{\textwidth}
    \centering
    \epsfig{file = 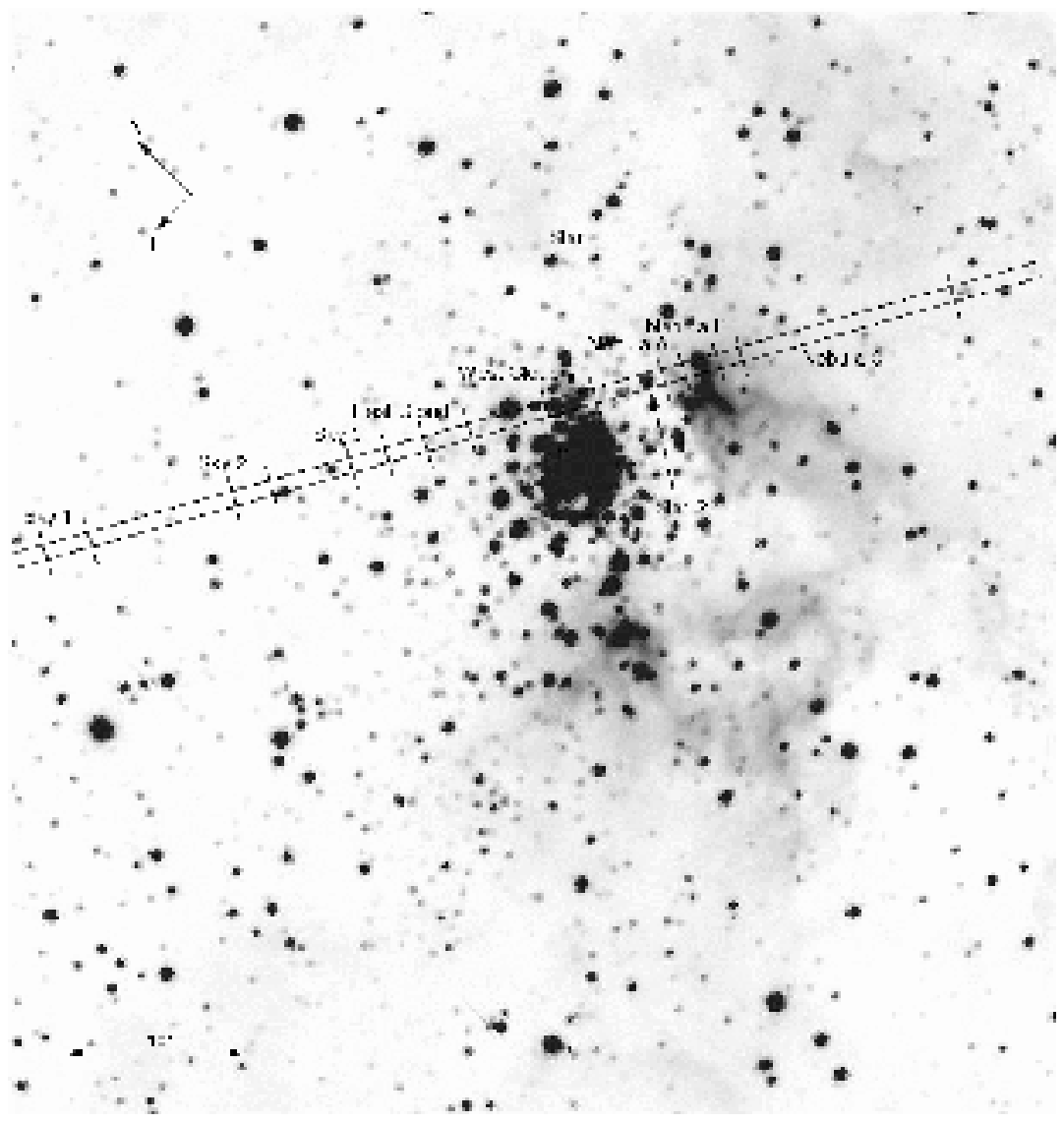, width = 103mm}
  \end{minipage}\\[10pt]%
  \begin{minipage}{\textwidth}
    \centering
    \epsfig{file = 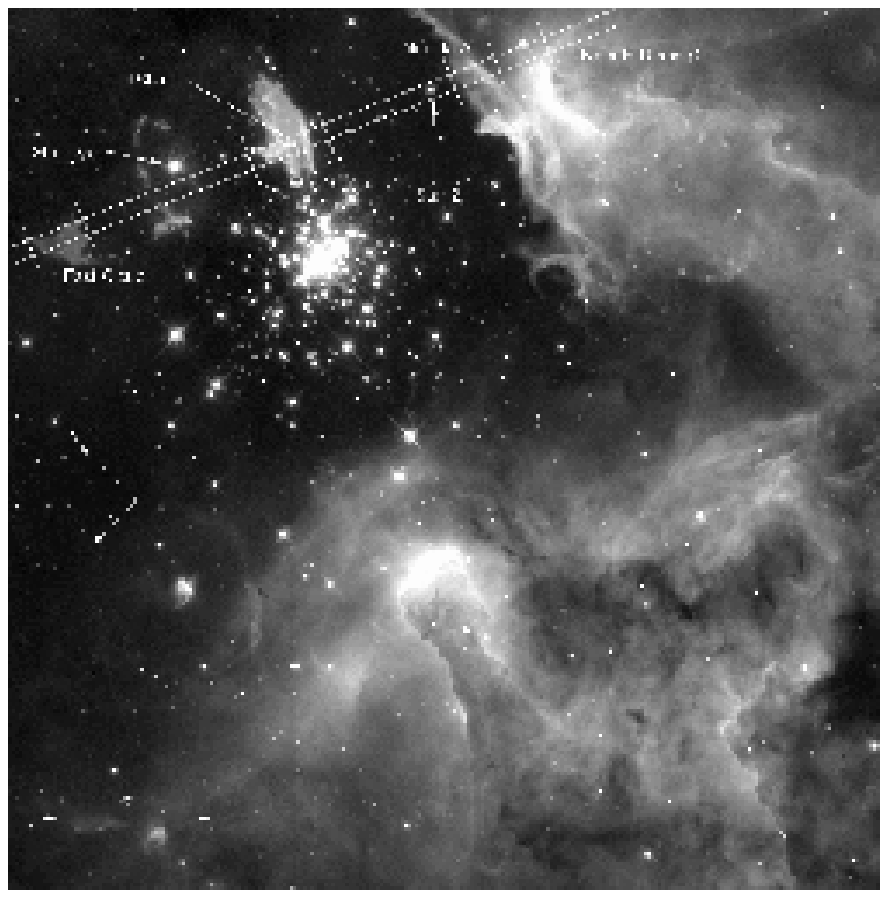, width = 103mm}
  \end{minipage}
  \caption[EW slit position showing the approximate width of the slit and apertures.]{EW slit position showing the approximate width of the slit and apertures where the 1D spectra were extracted. The top finder is a 10\,s $R$-band image taken with EMMI on the NTT at La Silla, Chile. The bottom is an $HST$ image of Sher\,25 and surrounding nebula showing the EW slit position and apertures in more detail. The width of the slit does not reflect the difference in resolution of the NTT and $HST$ images.}
  \label{fig:EW}
\end{figure*}

\begin{figure*}
  \begin{minipage}{\textwidth}
    \centering
    \epsfig{file = 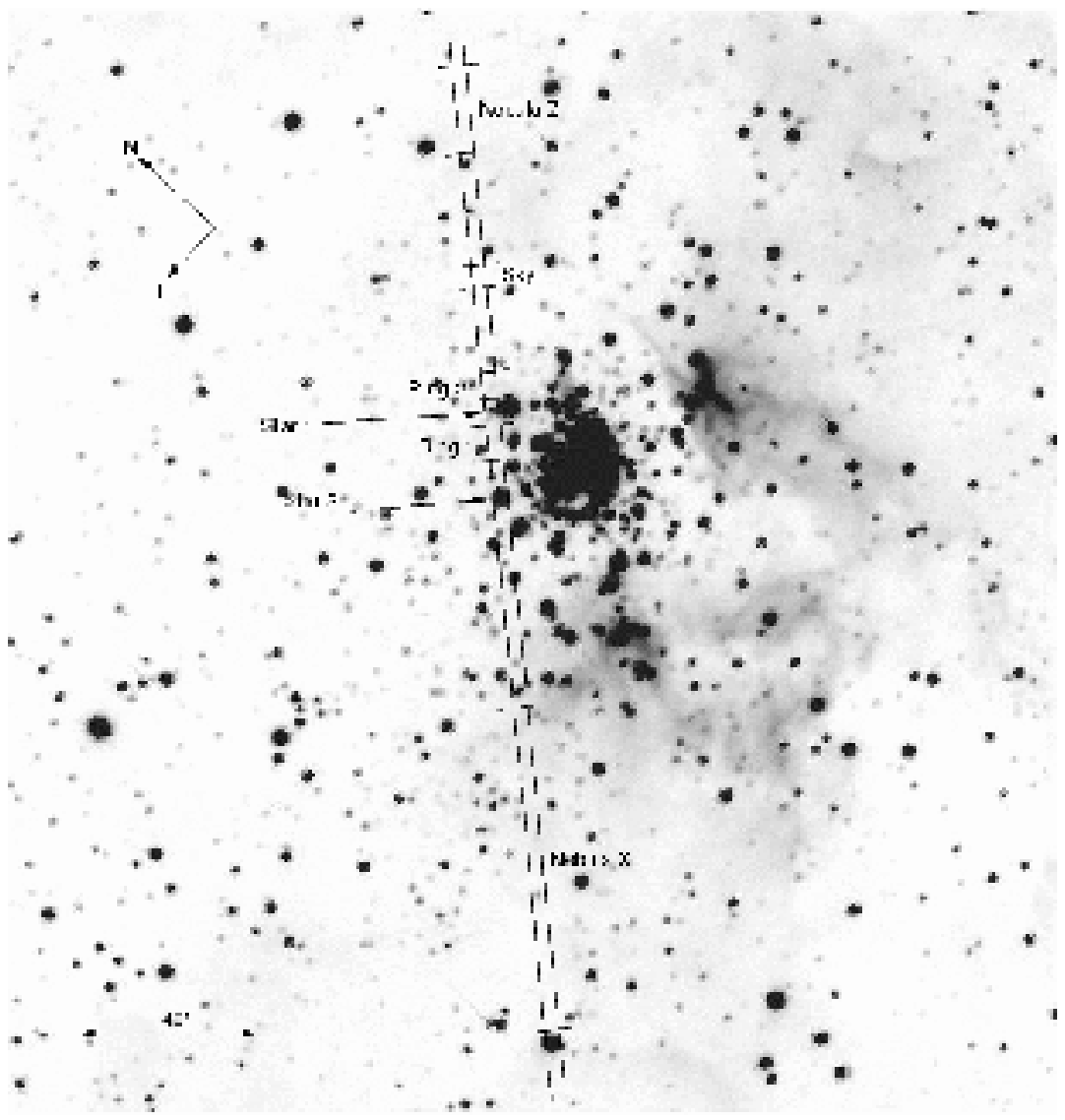, width = 103mm}
  \end{minipage}\\[10pt]%
  \begin{minipage}{\textwidth}
    \centering
    \epsfig{file = 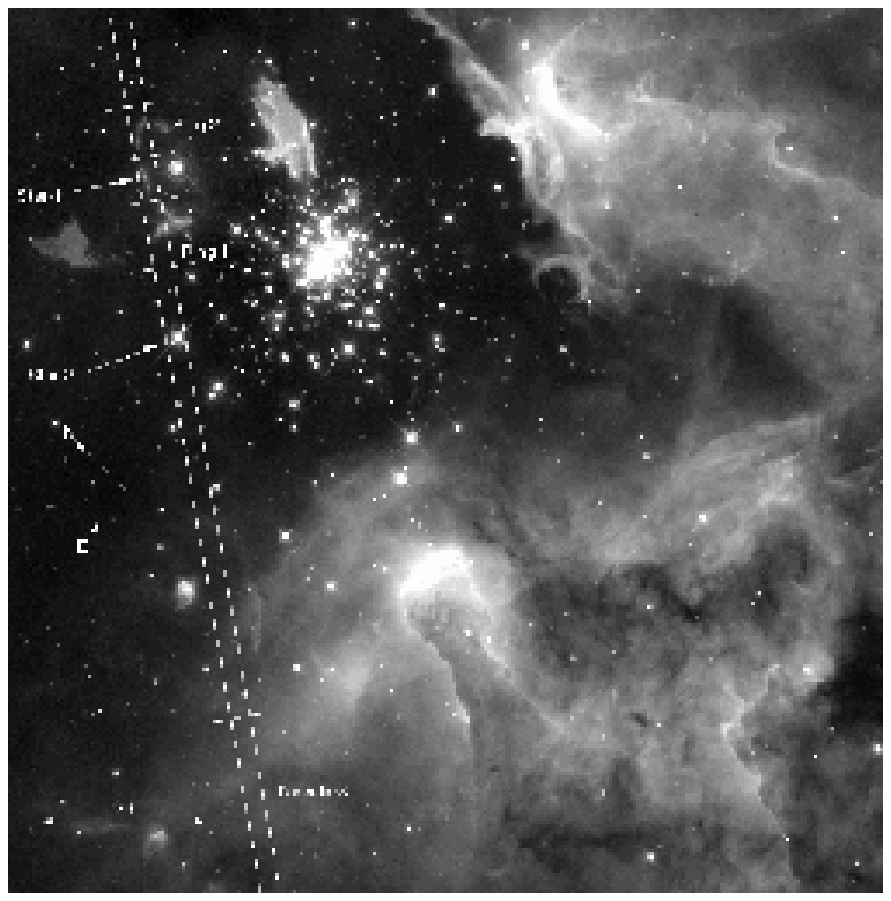, width = 103mm}
  \end{minipage}
  \caption[NS slit position showing the approximate width of the slit and apertures.]{NS slit position showing the approximate width of the slit and apertures where the 1D spectra were extracted. The images are the same as in Figure~\ref{fig:EW}. Again the $HST$ finder shows the NS slit position and apertures in more detail and the width of the slit does not reflect the difference in resolution of the NTT and $HST$ images.}
  \label{fig:NS}
\end{figure*}

Long-slit spectroscopy of the nebula was obtained to enable
a detailed abundance analysis of the two bipolar lobes, the ring
of Sher\,25 and the background nebula of NGC\,3603. The observations
were taken with the ESO 3.6-m telescope at La Silla, Chile with the
ESO Faint Object Spectrograph and Camera (EFOSC2) on two consecutive
nights commencing on 2003 May 7. Exposures were taken at two slit positions using
Grism \#11 with a 1$''$ slit.  The slit on the first night was orientated
approximately east-west (EW), observing the bipolar lobes, while on the second
night the slit was approximately north-south (NS), observing the ring. A
list of the observations is given in Table~\ref{tab:neb}. The slits
were aligned using stars 1 and 2, labelled in the finders
in Figures~\ref{fig:EW} and \ref{fig:NS}. The finders also show the
apertures extracted along the different slit positions. The
ground-based finders, shown in the top figures, are from a 10\,s
$R$-band image taken with the ESO Multi Mode Instrument (EMMI) on the
ESO 3.58-m New Technology Telescope (NTT) at La Silla, Chile. The
bottom figures are $HST$ finders of Sher\,25 and the surrounding nebula
showing the slit positions and the extracted apertures in more
detail. The width of the slit in the figures does not reflect the difference in
resolution of the NTT to the $HST$ images. 
The final resolution of the spectra, as measured from the FWHM of the night sky lines 
was $\sim$14\,\AA, with spectral coverage from 3550 to 7345\,\AA.

\begin{table}
  \caption{Journal of spectroscopic observations of the Sher\,25
  nebula. The usable wavelength range of each spectrum was
  3550--7345\,\AA, at a resolution of $\sim$14\,\AA.\label{tab:neb}}
  \vspace{3mm}
  \small
  \centering
  \begin{tabular}{lrrrrlll}
    \hline
    \hline\\[-11pt]
    Date & JD & PA & Airmass & Set\\
    & (245 0000+) & (degree) & & \\[2pt]
    \hline\\[-11pt]
    2003 May 7 & 2767.49 & $-22.55$ & 1.20 & 1st\\
    2003 May 7 & 2767.51 & $-10.41$ & 1.18 & 1st\\
    2003 May 7 & 2767.53 & 2.11 & 1.18 & 1st\\
    2003 May 7 & 2767.56 &  17.01 & 1.19 & 2nd\\
    2003 May 7 & 2767.58 &  29.27 & 1.21 & 2nd\\
    2003 May 7 & 2767.60 &  40.25 & 1.24 & 2nd\\
    2003 May 8 & 2768.48 & $-28.46$ & 1.21 & 1st\\
    2003 May 8 & 2768.50 & $-16.62$ & 1.19 & 1st\\
    2003 May 8 & 2768.52 & $-4.25$ & 1.18 & 1st\\
    2003 May 8 & 2768.55 & 11.44 & 1.18 & 2nd\\
    2003 May 8 & 2768.57 & 23.55 & 1.20 & 2nd\\
    2003 May 8 & 2768.59 & 34.94 & 1.23 & 2nd\\
    \hline
  \end{tabular}
\normalsize
\end{table}

The spectra were reduced using standard techniques within {\sc
iraf}. The frames were debiased, then flat-fielded using normalised dome
flats. Six target spectra, each 1800\,s long,
were obtained on both nights, with He-Ar arcs taken before, after and during
the run. Unfortunately a small drift occurred during the arc exposure
in the middle of the run so the processed target spectra were combined
into two different sets for each night; before and after the arc (1st
and 2nd sets in Table~\ref{tab:neb}). The different sets were then
calibrated separately and later the wavelength calibrated, extracted
1D spectra were combined. 

Long-slit spectra suffer from geometric
distortions introduced by a number of factors including differential
atmospheric diffraction, the camera optics and the grism. These
distortions were removed from the 2D spectra by a geometric
transformation of the standards, combined arcs and the
target spectra using {\sc iraf} tasks within the {\sc longslit}
package in {\sc twodspec}. Using the {\sc identify} and {\sc
reidentify} tasks spectral lines were identified in the arc spectra
along the dispersion axis and reidentified at a constant interval over
the whole 2D spectrum. The {\sc fitcoords} task was then used to fit a
2D function to the wavelength with respect to the pixel coordinate,
producing a wavelength transformation map. The geometric correction
was then applied to the arc using the {\sc transform} task. The
wavelength map was checked by comparing the corrected arc to the
original before it was applied to the standards and target spectra.

In both sets the transformation of the EW slit was not well determined
in the blue near \Oii\ 3727, as the spectral feature identified in the
calibration arc was quite weak and unreliable. The wavelength solution
in the blue was therefore slightly uncertain. No such problem,
however, was encountered during the transformation of the NS
slit. Even with this problem, the transformation greatly improved the
distortions in the spectra for both slit positions. The 2D
wavelength-calibrated target spectra were then flux calibrated using
the geometrically-corrected standard spectra (extracted
using {\sc apall}). Only standard star observations that were observed
at a low airmass and with a larger slit-width were used for flux calibration.
In both slit positions at least two standards were used. The spectra
were corrected for extinction and flux calibrated using the standard
{\sc iraf} tasks {\sc standard, sensfunc} and {\sc calibrate}. The 2D
spectra could not be background subtracted as the transformation at
the blue end was not perfect. This was because the calibration arc had
no suitable lines in the far blue, causing a small distortion to
remain. The sky was instead subtracted from the extracted 1D spectra.

\begin{figure}
    \centering
    \epsfig{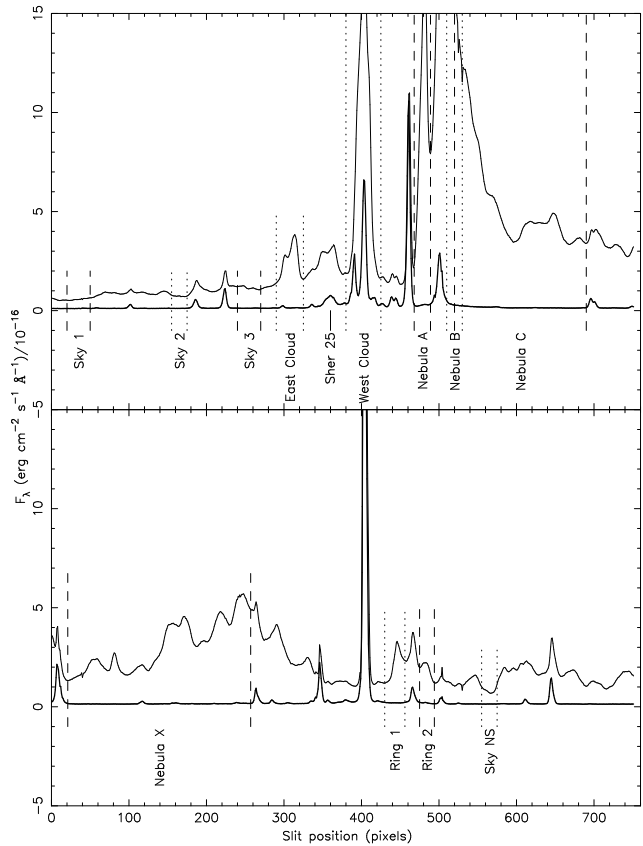}
  \caption[\ha\ emission and the stellar continuum as a function of
  slit position for the EW and NS slit position.]{H$\alpha$ emission
  and the stellar continuum (bold line) as a function of slit position
  for the EW (upper panel) and NS (lower) slits. The apertures marked are the
  exact apertures that were extracted and correspond to those
  indicated in Figures~\ref{fig:EW} and \ref{fig:NS}. The pixel scale
  is approximately $\sim$0.4$''$.}\label{fig:slit}
\label{fig:ap}
\end{figure}

The extraction apertures for the target and sky spectra, shown in
Figures~\ref{fig:EW} and \ref{fig:NS}, were identified by comparing
the \ha\ emission to the stellar continuum along the slit. The exact
apertures that were extracted are shown in Figure~\ref{fig:ap} and
correspond to those in Figures~\ref{fig:EW} and \ref{fig:NS}. The sky
spectra in the EW slit were extracted from the east end of the slit
where the \ha\ emission was at its lowest (Figure~\ref{fig:ap}, upper
panel). The sky from the EW slit was combined into two different
spectra, Sky$_{123}$ and Sky$_{12}$, to get a better handle on the
errors associated with the sky-subtraction. Sky$_{123}$ and Sky$_{12}$
were found to differ slightly in the nebular lines, particularly in
the Balmer, N and O lines. The night sky lines and background features
were the same in both spectra. As can been seen in Figure~\ref{fig:ap}
(lower panel) the background nebulosity extends over the full length
of the NS slit, so the sky spectrum was extracted from the region
where the \ha\ emission dips to its lowest level.  The NS and EW skies
were quite similar, as we would desire, with Sky$_{123}$ in better
agreement agreement with the NS sky spectrum. The 1D target spectra
were extracted using the {\sc apall} task from the wavelength and
flux-calibrated 2D spectra using the brightest star as a template. The
1D spectra were then read into the spectral analysis program {\sc
dipso} \citep{SUN50.24} for sky subtraction and further analysis.

As can be seen from the images of NGC\,3603 and Figure~\ref{fig:ap},
the background is highly complex and we selected the background
regions for subtraction from the Sher\,25 nebula as carefully as
possible. Ideally it would be desirable to sample the background
adjacent to the nebular extraction window. However, at our
ground-based resolution this would have led to the background often
being contaminated with stellar continuum, or high regions of
background flux that were not necessarily representative of the
background in the nebular aperture. This is perhaps the major
limitation of this ground-based study; spectroscopy at the resolution
of the {\it HST} images would undoubtedly improve this. Thus, we
demonstrate throughout this paper the consequence of using two different
sky selections, and propagate that through the abundance analyses.

\begin{figure*}
\centering
\epsfig{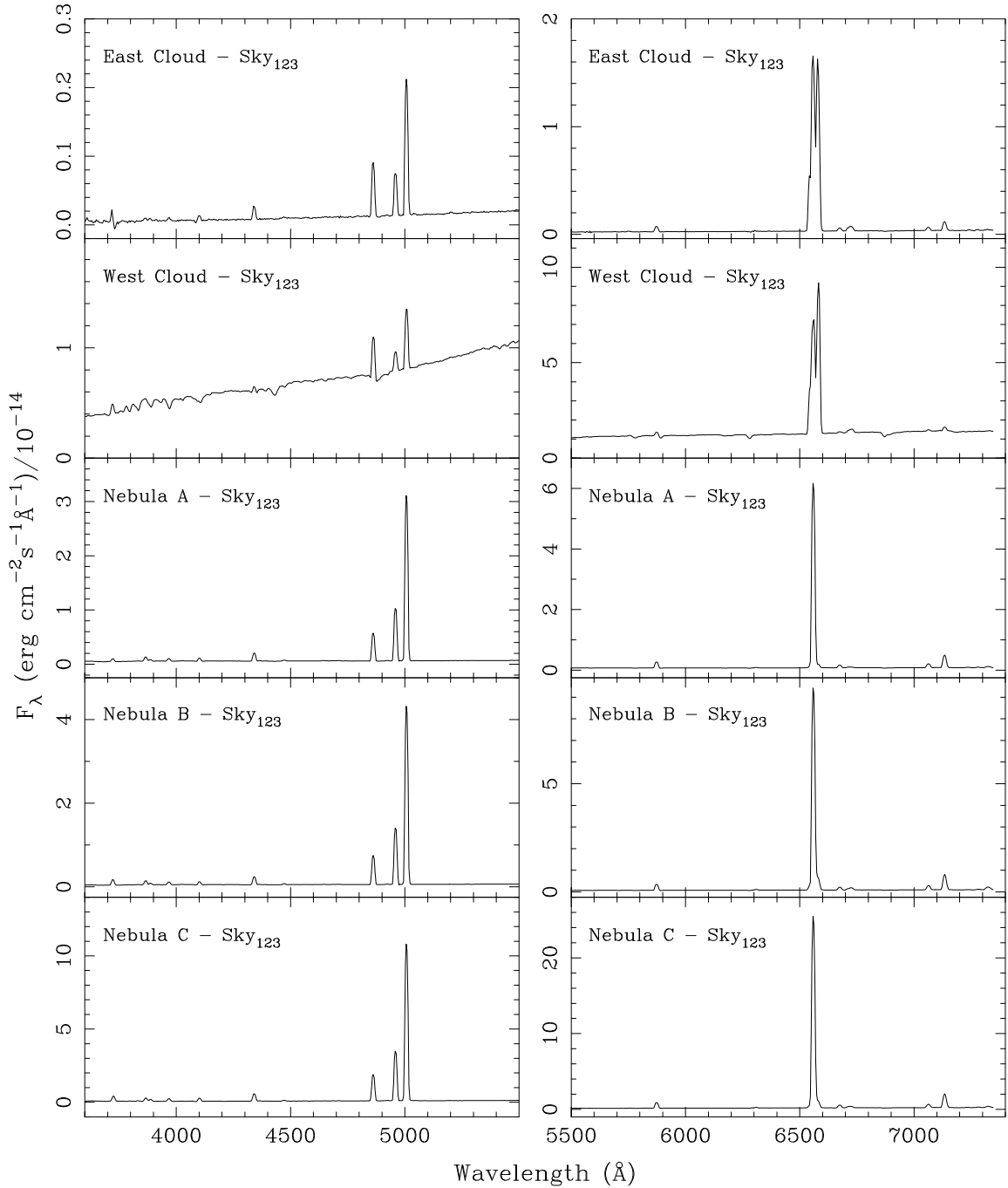}
\caption{Mosaic of the spectra from the EW slit position, in which the sky subtraction used all three extraction
regions, i.e. `Sky$_{\rm 123}$'.}\label{fig:EWspec}
\end{figure*}

\begin{figure*}
\centering
\epsfig{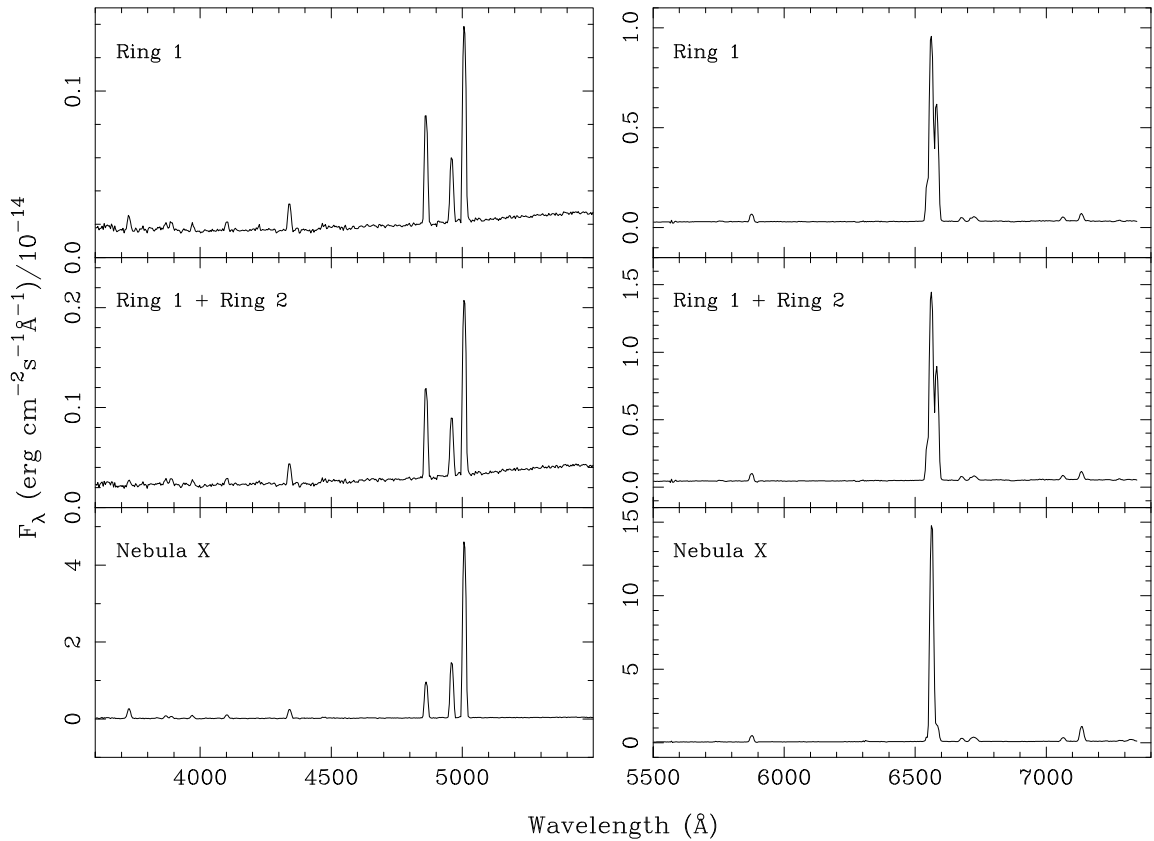}
\caption{Mosaic of the spectra extracted from the NS slit position.}\label{fig:NSspec}
\end{figure*}

\begin{figure*}
  \centering
  \epsfig{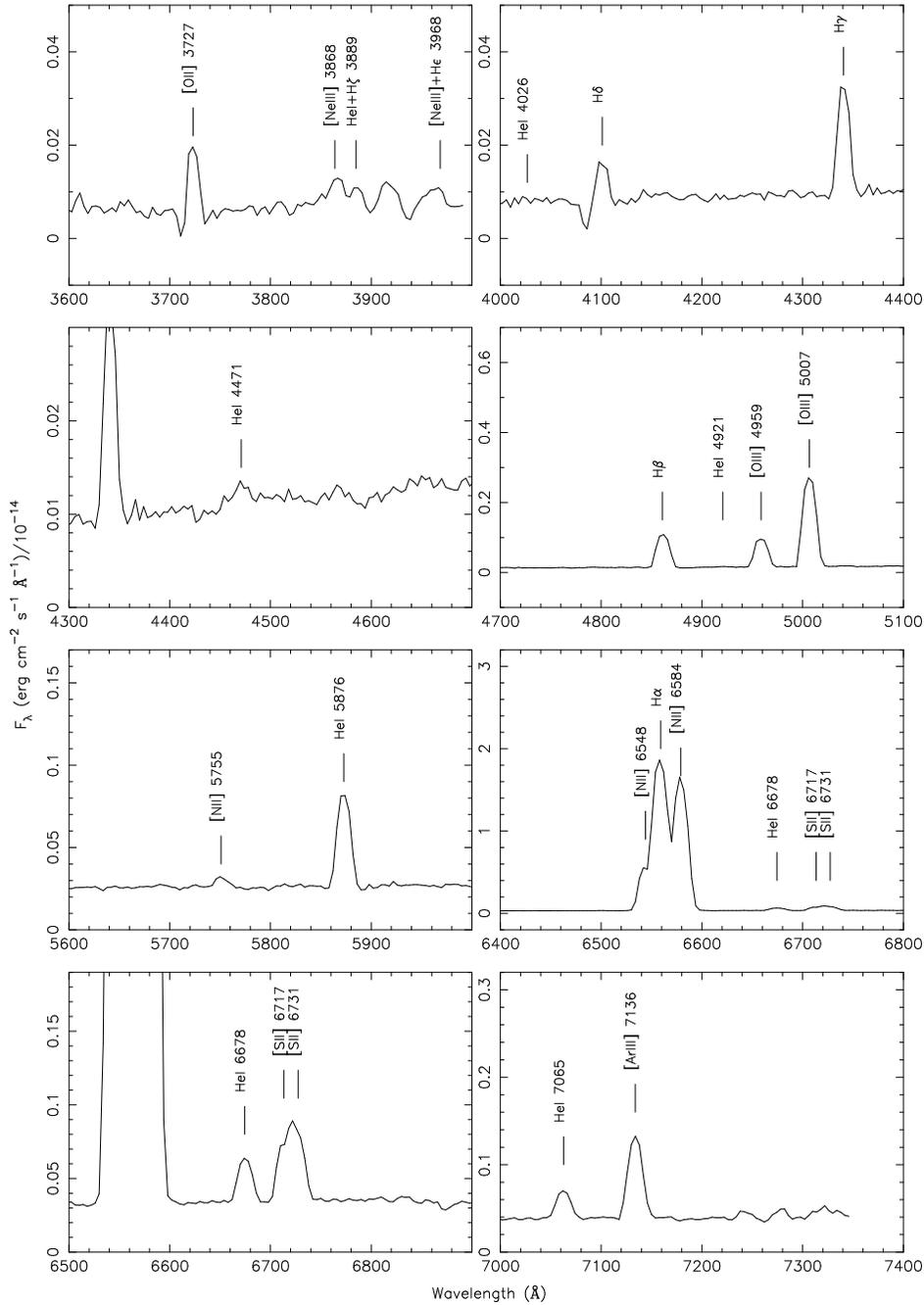}
  \caption[]{The spectrum of the East Cloud with line identifications.}
\label{fig:EC}
\end{figure*}

The extracted 1D spectra were sky subtracted, then the first and second
sets of both the EW and NS target spectra (Table~\ref{tab:neb}) were
combined within {\sc dipso}. The EW spectra were then sky subtracted
and co-added. The night sky lines were fully subtracted in all the
spectra illustrating the success of the sky subtraction. The
extracted, sky-subtracted target spectra are shown in
Figures~\ref{fig:EWspec} and \ref{fig:NSspec}, and a line
identification is given in Figure~\ref{fig:EC}. The line fluxes were
measured in {\sc dipso} using the emission line fitting package {\sc
elf}. The errors in the line fluxes were estimated from the
root-mean-square (RMS) of the continuum and were taken to be $2 {\it
FWHM} \times {\it RMS}$, where the $2 {\it FWHM}$ represents the width
of the line where the flux is 6\% of the peak, i.e. the width at the
base of the line. A 1\% flux calibration error was also combined in
quadrature.

\subsection{Stellar observations}
\label{sec:stellarobs}
Sher\,25 was observed using the University College London \'{E}chelle
Spectrograph (UCLES) on the Anglo-Australian Telescope (AAT) on the
night of 1999 May 28, as described in Paper~I. Two wavelength regions
were observed, one centred at 4335\,\AA, with the other at 6560\,\AA.
The UCLES spectra lacked good temporal coverage, but the blue-region
exposures were still check for evidence of a companion in Paper~I,
with no velocity shifts found at a significance greater than the
estimated error of $\sim$6~\kms.  Sher\,25 has since been observed
using the Fibre-fed, Extended Range Optical Spectrograph (FEROS) on
the 2.2-m Max Planck Gesellschaft (MPG)/ESO telescope on 2004 July
5--7 and again six months later with the same instrument.  FEROS is a
fixed configuration instrument, with $R$ = 48,000, giving a wide
wavelength coverage of 3600--9200\,\AA\ in one exposure. The data were
reduced using the standard FEROS data reduction pipeline, which runs
under the {\sc midas} environment. A journal of these observations can
be found in Table~\ref{tab:spec}.

\begin{table*}
  \caption[Journal of spectroscopic stellar observations of Sher\,25.]{Journal of spectroscopic stellar observations of Sher\,25.}
  \centering
  \begin{tabular}{lrrrrlll}
    \hline
    \hline\\[-11pt]
    Date & JD & Range & Resolution & Telescope + & Observer\\
    & (245 0000+) & (\AA) &  (\AA) & Instrument &\\[2pt]
    \hline\\[-11pt]
    1999 May 28 & 1326.95 & 6534--6589 & 0.17 & AAT+UCLES & Smartt\\
                & 1327.00 & 3914--5284 & 0.11 & & \\
    2004 Jul 05 & 3191.60 & 3796--5030 & 0.09 & MPG/ESO+FEROS & Evans/Hendry\\
                &         & 6346--6783 & 0.14 &               &\\
    2004 Jul 07 & 3193.56 & 3800--5007 & 0.09 & MPG/ESO+FEROS & Evans/Hendry\\
                &         & 6346--6783 & 0.14 &               &\\
    2004 Dec 30 & 3369.85 & 3802--5008 & 0.09 & MPG/ESO+FEROS & Burnley\\
                &         & 6345--6779 & 0.14 &               &\\
    2004 Dec 31 & 3370.85 & 3796--5003 & 0.09 & MPG/ESO+FEROS & Burnley\\
                &         & 6345--6779 & 0.14 &               &\\
    2005 Jan 01 & 3371.83 & 3804--5008 & 0.09 & MPG/ESO+FEROS & Burnley\\
                &         & 6345--6778 & 0.14 &               &\\
    \hline
  \end{tabular}\\[3pt]
  \scriptsize
  AAT~=~4-m Anglo-Australian Telescope, Australia. MPG/ESO~=~2.2-m Max Planck Gesellschaft/ESO telescope, La Silla, Chile.\label{tab:spec}
  \normalsize
\end{table*}

\section{Nebular and stellar abundance analysis}\label{sec:neb}

\subsection{Nebular abundance method}

The nebular abundances were derived using the `direct' method
described by \citet{1998salg.conf..457S}. The logarithmic extinction
at \hb, $C(\hb)$, was derived from the \ha/\hb, \hg/\hb\ and \hd/\hb\
ratios, whilst simultaneously solving for the effects of the underlying
stellar absorption. The equivalent width of the stellar absorption,
$W_{\rm abs}$, was assumed to be equal for all four Balmer lines. The
theoretical Balmer line ratios were calculated using the H
emissivity calculations from \citet{1987MNRAS.224..801H}, which are
very similar to the values calculated by
\citet{1971MNRAS.153..471B}. The Galactic reddening law used was that
of \citet{1989ApJ...345..245C}, assuming $R_V = 3.1$. The extinction,
stellar absorption and their associated errors were determined from
Monte Carlo simulations, using the method described by
\citet{2001NewA....6..119O}. In some instances the \hd\ line was not
present, in which case its flux was set to a negligible amount. The
density assumed throughout this analysis was $100\,{\rm cm}^{-3}$,
because of the lack of resolved lines that can be used as density
diagnostics. In the low-density limit, which is defined as being
significantly lower than the critical densities, the emissivities do
not vary appreciably with density for any of the transitions involved.

A reliable measurement of the electron temperature of the ionised gas
is necessary for the `direct' conversion of emission line strengths
into ionic abundances. The electron temperatures were measured in most
cases using the \Nii\ $\lambda 5755/\lambda 6584$ and the \Oiii\
$\lambda 4363/(\lambda 4959+\lambda 5007)$ line ratios. The
temperature was found from an analytical fit to these line ratios
calculated for different temperatures using the {\sc ionic} task
within the {\sc iraf nebular} suite of programs. This suite of
programs is based on the {\sc fivel} program developed by
\citet{1987JRASC..81..195D} and described by
\citet{1995PASP..107..896S}. Wherever possible the N$^+$ and the
O$^{++}$ electron temperatures, $T_{\rm e}(\rm N^{+})$ and $T_{\rm
e}(\rm O^{++})$, were measured although it was not possible to measure
$T_{\rm e}(\rm O^+)$ because of the wavelength range of the spectra.

A single electron temperature cannot be adopted for the whole
nebula. \citet{1980A&A....85..359S} showed that in high metallicity
nebulae the temperature increases with radius. This is because of the
more efficient cooling of the \Oiii\ fine-structure lines in the inner
parts of the nebula, where O$^{++}$ is more dominant. O$^{+}$, which
is more abundant at larger radii, is not as efficient a coolant as
O$^{++}$, because it does not have similar IR fine-structure lines. Oxygen
ions play a dominant role in the cooling process so, to a first
approximation, it is reasonable to treat the nebula as having two
different temperature zones, roughly corresponding to the O$^+$ and
the O$^{++}$ zones. In this model, $T_{\rm e}(\rm O^{++})$ is taken to
represent the temperature of the high ionisation species and $T_{\rm
e}(\rm O^+)$, or $T_{\rm e}(\rm N^+)$ in this case, the low-ionisation
species \citep{1992AJ....103.1330G}. However, photoionisation models
suggest that some ions, such as S$^{++}$ and Ar$^{++}$, do not fit into the
two-zone model and instead a three-zone model is more
appropriate. \citet{1992AJ....103.1330G} showed that the region where
S$^{++}$ predominates straddles the zones in which O$^{++}$ and
O$^{+++}$ dominate. The more appropriate estimation of $T_{\rm
e}(\rm S^{++})$ and $T_{\rm e}({\rm Ar}^{++})$ is given in
equation~(\ref{equ:S++}). This three-zone model is assumed here and
the \citet{1992AJ....103.1330G} relationships, which were adopted for
the other ionic temperatures, are given in
equations~(\ref{equ:S++})--(\ref{equ:N++}), where $T_{\rm e}$ is the
electron temperature in units of $10^4$\,K.
\begin{eqnarray}
T_{\rm e}({\rm S}^{++}) = T_{\rm e}({\rm Ar}^{++}) &=& 0.83 T_{\rm e}({\rm O}^{++})+0.17 \label{equ:S++}\\
T_{\rm e}({\rm O}^{+}) &=& 0.7T_{\rm e}({\rm O}^{++})+0.3 \label{equ:O+}\\
T_{\rm e}({\rm O}^{+}) = T_{\rm e}({\rm S}^{+}) &=& T_{\rm e}({\rm N}^{+})\label{equ:N+}\\
T_{\rm e}({\rm Ne}^{++}) &=& T_{\rm e}({\rm O}^{++})\label{equ:N++}
\end{eqnarray}
When $T_{\rm e}({\rm O}^{++})$ could not be measured, it was estimated
from equation~(\ref{equ:O+}), using $T_{\rm e}({\rm N}^+)$. In the few
cases where neither $T_{\rm e}({\rm N}^+)$ or $T_{\rm e}({\rm
O}^{++})$ could be measured, they were estimated from the simple
average of the other Sher\,25 or NGC\,3603 background spectra, with an
adopted error of 2,000\,K.

Ionic abundance ratios were then calculated from the reddening and
Balmer, absorption-corrected, emission line ratios, listed in
Table~\ref{tab:fluxes}, using equation~(\ref{equ:ionic}) where
$\epsilon$ is the theoretical emissivity and $N$ is the ionic
abundance by number of ion X$^i$.
\begin{eqnarray}
\frac{N({\rm X}^i)}{N(\rm H^+)} &=& \frac{I(\lambda)}{I(\hb)}\frac{\epsilon(\hb)}{\epsilon(\lambda)}\label{equ:ionic}\\
y^{++} = \frac{N({\rm He}^{++})}{N({\rm H}^+)} &=& 0.0816 \frac{F(\lambda4868)}{F(\hb)} T_{\rm e}({\rm O}^{+})^{0.145}\label{equ:y++}\\
\frac{N({\rm O}^{+++})}{N(\rm O)} &=& \frac{y^{++}}{0.081+y^{++}}\label{equ:O4}
\end{eqnarray}

 The theoretical emissivities were calculated from parabolic fits,
with respect to temperature, of the output from the {\sc nebular} task
{\sc ionic}.  The He$^{++}$ abundance is calculated from the \Heii\
$\lambda4686$ line using the relationship given in
equation~(\ref{equ:y++}), which is a fit to the coefficients of
\citet{1987MNRAS.224..801H}, and hence the O$^{+++}$ abundance is
estimated using equation~(\ref{equ:O4}).  The elemental abundances were
calculated from the sum of the ionic abundances from all the relevant
ionic states. When this is not possible,
the elemental abundance ratio was calculated from the available
ionic abundances with an ionisation correction factor (ICF) applied
instead. The ICFs used are those of \citet{1998ApJ...500..188I}, which
were found analytically from a fit to the model calculations of
photoionised \Hii\ regions by \citet{1990A&AS...83..501S}. They are
given in equations~(\ref{equ:ICF1})--(\ref{equ:ICF2}), where $x ={\rm
O}^+/{\rm O}$.

\begin{eqnarray}
ICF(\rm S) & = & \frac{S}{S^+ + S^{++}} \nonumber\\
       & = & (0.013+x\{5.10+x[-12.78+\nonumber\\&&\hspace{5mm}x(14.77-6.11x)]\})^{-1}\label{equ:ICF1}\\
ICF(\rm Ar) &=& \frac{Ar}{Ar^{++}+Ar^{+++}}\nonumber\\
            &=& \{0.99 + x[0.091 + x(0.077x - 1.14)]\}^{-1}\\
ICF(\rm Ar)& = & \frac{Ar}{Ar^{++}} \nonumber\\
       & = & [0.15 + x(2.39 - 2.64x)]^{-1}\label{equ:ICF2}
\end{eqnarray}

The measured line fluxes and equivalent widths are listed in
Table~\ref{tab:fluxes}, alongside the reddening and
absorption-corrected intrinsic line fluxes. Also listed in the table
are the \hb\ flux and the final reddening and underlying Balmer
absorption equivalent width corrections, $C(\hb)$ and $W_{\rm abs}$,
that were applied from the Monte Carlo
algorithm. Table~\ref{tab:NII-Ha} shows the flux ratios of \Nii\
$\lambda\lambda 6548$ and 6584 with respect to \ha, compared to those
of Sher\,25 and SN\,1987A from \citet{1997ApJ...475L..45B}.

The ionic and total abundances for the Sher\,25 nebula and NGC\,3603
background nebula are listed in Tables~\ref{tab:Sher25} and
\ref{tab:n3603}, respectively. The temperatures that were used to
derive these abundances are also listed. A summary of the total
abundance ratios of N/O and the total logarithmic abundance ratios of
O, N, S, Ne and Ar is given in Table~\ref{tab:summary} for both the
Sher\,25 nebula and the NGC\,3603 background nebula. The abundances
for the different background subtractions are given for the EW
spectra, with their averages in the following entries.  The average EW
abundances were then averaged with the NS abundances to give overall
values for the Sher\,25 and background nebulae, as given in bold font
in Table~\ref{tab:summary}.  The abundances from R1 (surrounded by
square brackets in the table) were not included in the averages.  The
Sher\,25 stellar abundances (from Section~\ref{sec:stellar}) are also
listed for comparison, as are solar abundances from
\citet{2004A&A...417..751A} and
\citet{2005ASPC..336...25A}. The solar Ar and Ne abundances, however,
should be treated with caution as they are indirect
measurements.

The \Nii/\ha\ line ratios for the ring from
\citet{1997ApJ...475L..45B} are consistent with the values measured
here (see Table\,\ref{tab:NII-Ha}).
However, \citeauthor{1997ApJ...475L..45B} find higher values for
both the East Cloud (or the poles in Brandner et al's nomenclature)
and the background nebula. The
\citeauthor{1997ApJ...475L..45B} spectra were background subtracted by
linear interpolation between the nebular emission to either side of
the extracted spectrum. This is where the difference probably enters,
as it was found in the spectra presented here that the emission close
to the extracted spectra was too similar to be background. It was
concluded that the object was quite extended and that the sky should
be extracted from further afield. In the further paper by
\citet{1997ApJ...489L.153B} the structure around Sher\,25 was shown to
be a full hourglass nebula, therefore it is possible that the
subtracted background was part of the structure and not only
background.

In general, the abundances for the three components of the Sher\,25
nebula (WC, EC and the ring) agree very well (see
Table~\ref{tab:summary}), with a maximum RMS scatter of 0.2\,dex
between the different spatial positions. The O abundances of the
Sher\,25 nebula are comparable with both those in the background
nebula and solar abundances, suggesting that the cluster and the
original composition of Sher\,25 was similar to solar.  The S
abundances of the background nebula agree well with the solar value
(within the errors), whereas the Ne abundances are significantly
higher than the solar and the background nebular abundances, although
the solar Ne abundance may be suspect
\citep{2005Natur.436..525D}.
The mean Ar abundances of the Sher\,25 nebula are also somewhat higher
than the background, although within the uncertainties they are not
incompatible.  The differences between the solar and nebular
abundances for Ne and Ar may arise not only from the indirect
measurements of the solar abundances, but also from the ICFs as they are
based on photoionisation models.  As we will discuss below, the
stellar O abundance from {\sc fastwind} is also in good agreement with
that from the nebula, within the expected uncertainties (see
Section~\ref{sec:stellar}). As discussed above in Section~\ref{sec:longslitobs},
there may be some contamination of background nebular flux in the
extracted Sher\,25 circumstellar apertures. As the main difference
between the surrounding nebula and that of Sher\,25 is the N
abundance, such contamination would work to reduce the measured
enhancement of N.  Hence, the very high N abundance appears to be a
robust result and is supported in the stellar analysis below.

 \begin{table*}
   \caption[\Nii/\ha\ line fluxes for Sher\,25 and
   Sk~\mbox{$-69\degree202$}/SN\,1987A.]{\Nii/\ha\ line fluxes for
   Sher\,25 and Sk~\mbox{$-69\degree202$}/SN\,1987A as listed by
   \citet{1997ApJ...475L..45B}. The poles of
   \citeauthor{1997ApJ...475L..45B} are directly comparable with the
   East Cloud of this paper.}\label{tab:NII-Ha}
   \centering
   \begin{tabular}{lrrr}
     \hline
     \hline\\[-11pt]
     Property & \multicolumn{2}{c}{Sher\,25} & Sk~\mbox{$-69\degree202$}/SN\,1987A\\
     & This Paper & \citet{1997ApJ...475L..45B} & \\[2pt]
     \hline\\[-10pt]
     (\Nii/\ha)$_{\rm ring}$  & $0.8:1$     & $0.9-1.2:1$ & $4.2:1^a$\\
     (\Nii/\ha)$_{\rm poles}$ & -- & $2.1:1$     & $2.5:1^a$\\
     (\Nii/\ha)$_{\rm EC}$    & $1.2-1.3:1$ & --          & -- \\
     (\Nii/\ha)$_{\rm WC}$    & $1.8:1$     & --          & -- \\
     (\Nii/\ha)$_{\rm background}$ & $0.02-0.07:1$ & $0.15:1$ & $0.09:1^b$\\
     \hline
   \end{tabular}\\[3pt]
   \scriptsize
   $^a$\citet{1996ApJ...459L..17P}; $^b$\citet{1997ApJ...475L..45B}\\
   \normalsize
 \end{table*}

\begin{table*}
\caption[]{Summary of the main elemental abundances of both
the Sher\,25 and background nebula (EC = East Cloud, WC = West Cloud,
R1 = Ring 1, R2 = Ring 2, NA = Nebula A, NB = Nebula B, NC = Nebula C
and NX = Nebula X).  The mean abundances from the
different sky subtractions are given as well as the average abundances
of the means for two different nebulae. R1, in the square brackets, is
not included in the mean abundance. 
Italics denote an indirect
measurement.}\label{tab:summary}
\centering
\begin{tabular}{lrrrrrr}
\hline
\hline\\[-11pt]
Object & N/O & $12 + \log({\rm O}/{\rm H})$ & $12 + \log({\rm N}/{\rm
H})$ & $12 + \log({\rm S}/{\rm H})$ & $12 + \log({\rm Ne}/{\rm H})$ &
$12 + \log({\rm Ar}/{\rm H})$\\[2pt]
\hline\\[-11pt]
EC$_{12}$   & 1.36(0.22) & 8.65(0.13) & 8.78(0.14) & -- & 8.42(0.15) & 6.70(0.13)\\
EC$_{123}$  & 3.07(0.97) & 8.63(0.13) & 9.12(0.17) & -- & 8.71(0.15) & 6.87(0.12)\\
EC (mean)$^a$   & 2.22(0.50) & 8.64(0.09) & 8.95(0.20) & -- & 8.57(0.18)  & 6.78(0.12)\\[4pt]
WC$_{12}$   & 1.50(0.22) & 8.64(0.18) & 8.82(0.19) & -- & --           & 6.53(0.19)\\
WC$_{123}$  & 1.54(0.22) & 8.62(0.17) & 8.81(0.18) & -- & --           & 6.49(0.19)\\
WC  (mean)$^a$  & 1.52(0.16) & 8.63(0.13) & 8.81(0.13) & -- & --           & 6.51(0.13)\\[4pt]
[R1         & 1.27(0.43) & 8.57(0.34) & 8.67(0.35) & -- & 8.29(0.36) & 6.65(0.33)]\\
R1+R2	    & 2.61(1.31) & 8.55(0.36) & 8.97(0.38) & -- & 8.35(0.38) & 6.81(0.33)\\[4pt]
{\bf Sher\,25 Nebula}$^b$ & {\bf 2.12(0.55)} & {\bf 8.61(0.13)} & {\bf 8.91(0.15)} & -- & {\bf 8.46(0.21)} & {\bf 6.70(0.17)}\\[10pt]
NA$_{12}$   & 0.07(0.01) & 8.60(0.07) & 7.46(0.07) & 7.29(0.05) & 8.17(0.07) & 6.44(0.06)\\
NA$_{123}$  & 0.07(0.01) & 8.59(0.11) & 7.44(0.12) & 7.28(0.06) & 8.17(0.12) & 6.44(0.08)\\
NA  (mean)$^a$  & 0.07(0.01) & 8.60(0.07) & 7.45(0.07) & 7.29(0.04) & 8.17(0.07) & 6.44(0.05)\\[4pt]
NB$_{12}$   & 0.12(0.01) & 8.56(0.08) & 7.66(0.08) & 7.52(0.04) & 8.15(0.08) & 6.42(0.07)\\
NB$_{123}$  & 0.14(0.01) & 8.64(0.08) & 7.79(0.09) & 7.46(0.05) & 8.23(0.09) & 6.46(0.07)\\
NB  (mean)$^a$  & 0.13(0.01) & 8.60(0.07) & 7.72(0.09) & 7.49(0.04) & 8.19(0.07) & 6.44(0.05)\\[4pt]
NC$_{12}$   & 0.08(0.02) & 8.46(0.09) & 7.35(0.13) & 7.09(0.05) & 8.00(0.09) & 6.33(0.06)\\
NC$_{123}$  & 0.06(0.02) & 8.49(0.10) & 7.30(0.14) & 7.09(0.05) & 8.04(0.10) & 6.35(0.06)\\
NC (mean)$^a$   & 0.07(0.01) & 8.47(0.07) & 7.32(0.10) & 7.09(0.04) & 8.02(0.07) & 6.34(0.04)\\[4pt]
NX	    & 0.07(0.01) & 8.55(0.23) & 7.39(0.24) & 7.01(0.25) & 8.04(0.24) & 6.54(0.24)\\[4pt]
{\bf Background Nebula}$^b$ & {\bf 0.09(0.03)} & {\bf 8.56(0.07)} & {\bf 7.47(0.18)} & {\bf 7.22(0.22)} & {\bf 8.11(0.09)} & {\bf 6.44(0.08)}\\[10pt]
Solar & 0.13(0.02) & 8.66(0.05)$^c$ & 7.78(0.06)$^d$ & 7.14(0.05)$^d$& {\it 7.84(0.06)}$^d$ & {\it 6.18(0.08)}$^d$\\
Sher\,25 {\sc fastwind} & 1.7$^{+0.7}_{-0.5}$ & 8.51(0.18) & 8.74(0.39) & -- & -- & -- \\
\hline
\end{tabular}
\flushleft\scriptsize
 Notes:  
$^a$ The errors for the averages are the errors from the individual abundances, combined with
an error for the sky subtraction (taken to be the difference
between the abundances and their average);
$^b$ The errors quoted for the overall abundances are either the standard deviation or the 
combined errors of the individual nebulae, which ever was the larger;
$^c$\citet{2004A&A...417..751A}; $^d$\citet{2005ASPC..336...25A}.
\normalsize
\end{table*}

\subsection{Stellar abundance analyses}
\label{sec:stellar}
A model atmosphere and abundance analysis of the stellar spectrum of
Sher\,25 was presented in Paper~I.  However, there
are now more sophisticated codes available that provide consistent,
unified treatments of the stellar wind and photospheric features.
Here we present results from re-analysis of the high-resolution UCLES
spectra using three contemporary model atmosphere codes; these
results now supercede those from Paper~I.
In the first instance we employed the photospheric code {\sc tlusty}
to determine atmospheric parameters and abundances in the
photosphere, assuming no wind contribution to any of the optical lines
\cite[for details of the model grids used 
see][]{1995ApJ...439..875H,2007A&A...466..277H}.  We then analysed the
spectra using the stellar-wind codes {\sc fastwind}
\citep{2005A&A...435..669P} and {\sc cmfgen}
\citep{1998ApJ...496..407H,2003ApJ...588.1039H}.
The application of these codes to studies of B-type supergiants is
described in detail by \cite{2004A&A...417..217T} and
\cite{2006A&A...446..279C}, respectively.

\begin{figure}
\centering
\epsfig{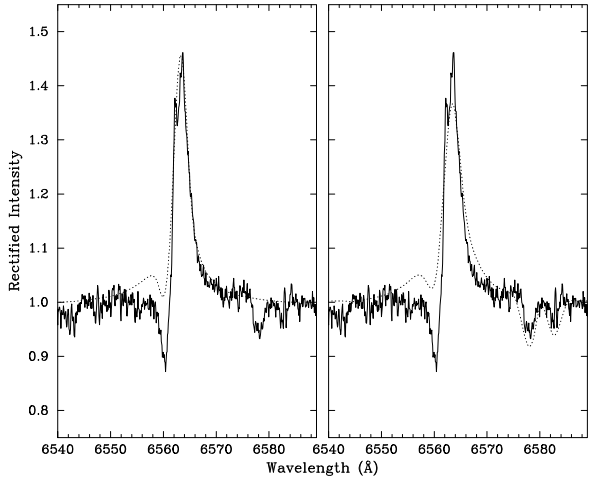}
\caption{{\sc fastwind} (left-hand panel) and {\sc cmfgen} (right) model
fits to the H$\alpha$ emission profile from the UCLES spectrum.  The C~{\scriptsize II} $\lambda\lambda$6578, 6583 
absorption lines
can be seen in the observed spectrum and the {\sc cmfgen} model, whereas the {\sc fastwind} model did not 
explicitly include carbon.}\label{Hafig}
\end{figure}

All three codes yield very similar physical parameters (summarised
in Table~\ref{stellar_params}), including similar mass-loss rates from the
two analyses of the H$\alpha$ profile.  Note that we adopted
slightly different values for $v_{\infty}$ for the {\sc fastwind} and
{\sc cmfgen} analyses -- both are reasonable values for early B-type
supergiants \citep{2004A&A...417..217T,2006A&A...446..279C}, and
changes in $v_{\infty}$ over the range of 750 to 1000~\kms\/ do not
affect the resulting abundances in either analysis.  Fits to the H$\alpha$
line are shown in Figure~\ref{Hafig}.

The nitrogen and carbon abundances are in reasonable agreement between
the new methods (within the typical uncertainty of $\pm$0.2\,dex),
however the absolute oxygen abundance is significantly lower in the
{\sc cmfgen} analysis than the other two.  As Sher\,25 is a very
luminous supergiant with a strong wind, we favour adopting the results
of either {\sc fastwind} or {\sc cmfgen} as the most reliable measure
of photospheric abundances.  We suggest that the {\sc fastwind}
abundances are the better estimate, as we employed all of the 15 O{\sc
ii} and 7 N{\sc ii} lines that were included in the observed spectra,
and for which reliable atomic data exists.  The {\sc cmfgen} analysis
used profile fitting to a smaller number of selected lines. For
example, for oxygen these were O{\sc ii}
$\lambda\lambda$4069-4092\,\AA\ and $\lambda\lambda$4590-96\,\AA. The
use of more lines in the {\sc fastwind} analysis, and
the resulting mean abundance, mitigates against systematic
uncertainties in atomic data for any particular multiplet. As we want
the best estimate of absolute abundances in the stellar photosphere
for comparison with the nebular results and solar-like abundances, we
therefore adopt the {\sc fastwind} values as the best comparison. The
uncertainties in the table are the standard error of the mean, and are
just a measure of the statistical scatter of the results from the
individual atomic transitions. As discussed by
\cite{2004A&A...417..217T} and
\cite{2007A&A...466..277H} the typical systematic uncertainty from
errors on the derivation of atmospheric parameters are around
$\pm$0.2\,dex. This should be remembered when comparing the stellar
results with the nebular abundances.

A new distance to NGC\,3603 has been calculated recently by
\cite{arXiv:0712.2621} and they compiled a list of all distances
quoted in the literature. Their favoured distance of 
7.6\,kpc and consistent treatment of line-of-sight extinction
leads them to suggest a luminosity of Sher\,25 of 
$\log L/L_{\rm \odot} = 5.78$\,dex. Their comparison with the
Geneva tracks \citep{1994A&AS..103...97M} would then suggest
an initial mass of $50\pm10$\msun. 

\begin{table*}
\caption{The atmospheric parameters derived for Sher\,25 from 
the three new analyses using the {\sc tlusty}, {\sc fastwind} and {\sc
cmfgen} model atmosphere codes.  The results from Paper~I
are also listed under NLTE\,H/He as this analysis did not include
metal-line blanketing in the model atmosphere and wind
computations. The uncertainties quoted are the standard error of the
mean, where the mean is from all lines of the element available (see
Paper~I for a full line listing). The {\sc cmfgen} results
assume $\log L/L_{\rm \odot}=5.80$.}
\label{stellar_params}
\vspace{ 0.1 cm}
\begin{tabular}{lllll}\hline\hline
 \vspace{ 0.1 cm}
Parameter                          & NLTE H/He &  {\sc tlusty}       & {\sc fastwind}      & {\sc cmfgen} \\ \hline
$T_{\rm eff}$   (K)                & $22300\pm1000$ & $21500\pm1000$  & $22000\pm1000$  & $21000\pm1000$        \\
$\log g$  (cgs)                    & $2.6\pm0.10$   & $2.6\pm0.20$    & $2.6\pm0.15$    & $2.5\pm0.20$         \\
$\dot{M}$ ($10^{-6}\msun\,yr^{-1}$)&  1.75          & ...             & 1.95	       & 1.60          	    \\
$v_{\infty}$ (\kms)         & 	     1000          & 	     ...            & 1000	       & 750                 \\
$\beta$                      & 	     1.5      & 	     ...          & 2.5	       & 2.5                  \\
$v_{turb}$ (\kms)            &       15       & 20              & 23              & 15           \\
C                            & 7.01$\pm$0.06 &       $7.82\pm0.16$   & ...	       & $7.8$                  \\
N                            & 8.42$\pm$0.12 &       $8.52\pm0.09$   & $8.74\pm0.15$   & $8.5$        \\
O                            & 8.87$\pm$0.07 &       $8.50\pm0.03$   & $8.51\pm0.05$   & $8.1$        \\
Mg                           & 7.46  	     &       $7.59\pm0.36$   & ...             & ...                  \\
Si                           & 7.42$\pm$0.07 &       $7.40\pm0.20$   & $7.61\pm0.08$   & ...        \\\hline
\hline
\end{tabular}
\end{table*}

\subsection{Discussion of abundances}

The most significant result from this analysis is the enhanced N
abundance of the Sher\,25 nebula compared to the background
nebula and solar abundances. The N abundance of the nebula is 0.17\,dex
larger than the stellar values from {\sc fastwind}.  However,
considering the uncertainties in the absolute values of both, the
difference is not significant and we consider the results to be
consistent. The N/O ratio (by number) in the nebula and the star are
also consistent within the uncertainties, and both are significantly
higher than that previously reported by Paper~I. The reason for the
lower value in Paper~I was the very high O abundance
derived. As discussed above, we believe that the {\sc fastwind} models
produce more reliable results than the previous spherical, blanketed
models. The fact that the new {\sc tlusty} results, {\sc fastwind}
abundances, and nebular values are in good agreement supports our
conclusion that the N/O of the stellar photosphere and nebula is in
the range $1.7-2.1$.

The morphology of the nebula and the strength of the N emission 
lines were suggested by Brandner et al. (1997a,b) to be a 
consequence of the star having gone through a previous RSG 
phase. A similar evolutionary path for Sk~\mbox{$-69\degree202$}
has been proposed to explain
the hourglass-shaped nebula now seen around SN\,1987A. The 
very high N/O which is quantitatively similar in 
both the stellar photosphere and the nebula could be 
interpreted as supporting this scenario. 
\citet{2001ApJ...551..764L} presented the
abundances that are predicted for the stellar envelope by the models of
\citet{1994A&AS..103...97M} after convective mixing during the RSG
phase. As discussed above, the best estimate for the initial 
mass of Sher\,25 is $50\pm10$\msun. 
Adopting a typical mass-loss rate of $2-5\times10^{-6}\,\msun\,{\rm y}^{-1}$
\citep{1996A&A...305..171P} and a main sequence (core hydrogen burning)
lifetime, $t_{\rm MS}$, of 4\,Myr, the star is likely to have lost
around 8--20\msun.  This would leave Sher\,25 with 60--85\,\% of its
mass at the onset of the convective
mixing. \citet{2001ApJ...551..764L} plotted the logarithmic N/O
abundance as a function of the remaining mass fraction for several
masses. The models indicate that Sher\,25 should have a N/O abundance
in the range of $2\lesssim{\rm N/O}\lesssim20$ if it had gone through
a previous RSG phase, and was originally a 40-60\msun\ main-sequence
star. Hence our N/O result is just consistent with the lower end of this
range, and one could interpret this as support for the star being in a
post-RSG phase. Similar N enhancements for stars on blue loops are
found by \cite{2000ApJ...544.1016H}, who find that a 25\msun\ star
should show an N enhancement of around 1.0\,dex after having passed
through RSG dredge-up.

However, we suggest that it is quite possible that the star
has not been through a RSG phase and that the
N enrichment we see is due to rotational mixing while the 
star was on the main-sequence. For example \citet{2001ApJ...551..764L}
also predict the surface abundances for rotationally-induced mixing as
a function of mixing time, and our abundances are more consistent
with this scenario.
The authors predict N/O abundances in the
range of $1\lesssim{\rm N/O}\lesssim40$ for weak-to-strong
mixing. Weak mixing has a mixing time greater than 5\,$t_{\rm MS}$, and
strong mixing less than 0.5\,$t_{\rm MS}$. The N/O abundance of the
Sher\,25 nebula  is consistent
with a moderate amount of mixing, with a mixing time of 2.5\,$t_{\rm
MS}$. The N
abundance is also greatly enriched by almost a factor of 30 above the
background nebula and around 13 times that of the solar levels. 
Similar values for N/O enrichment are also predicted by
\cite{2000ApJ...544.1016H}, although their most massive models were
for 25\,\msun. The post-RSG scenario has a further problem in that a
star of initial mass 50\,\msun\ is somewhat too massive to have been a
RSG with a convective envelope. Observationally we tend not to see
RSGs with luminosities that would place their initial masses above
25\,\msun\ \citep{2005ApJ...628..973L}. Theoretical tracks at
40\,\msun\ and above also do not reach the coolest regions where the
atmospheres are certain to become convective. Smith (2007) also points
out this problem with the RSG origin of the nebula for Sher\,25 citing
the lack of very massive, cool supergiants in observational H--R
diagrams of single stars in the Local Group
\cite[e.g.,][]{1994PASP..106.1025H}.  The consistency we see between
the nebular N and O abundances show that the nebula is of the same
composition as the stellar photosphere. We would suggest that this
likely due to an ejection event from Sher\,25 when it was in the BSG
phase rather than a combination of a RSG with a dense slow
wind being swept up by a subsequent faster wind from the hot B-type
phase \citep{1997ApJ...475L..45B}. Smith (2007) has also advanced
dynamical arguments that the newly discovered ring structure around
HD\,168625 was ejected in a LBV eruption while the star was in the hot
blue phase rather than being a swept up RSG wind. There seems to be a
body of evidence now arguing against the RSG origin for these ring
nebulae and perhaps an ejection mechanism while the star is still
hot. We note that HD\,168625 is of significantly lower luminosity and
lower initial mass ($\sim$20\msun) than Sher\,25 ($\sim$50\msun), and
it is interesting that such similar ring structures are found around
blue supergiants of vastly different mass. A mechanism for ejection of
a nebula which has bipolar lobes and a disk structure from a near
critically-rotating BSG has been proposed by
\cite{2007ApJ...666..967S}. This requires a pre-RSG phase and is
attractive for very massive stars with circumstellar nebulae showing
this type of morphology. Clearly Sher\,25 is not currently rotating
near its break-up velocity; its photospheric line-widths were given by
Paper~I as 98~\kms. This is an upper limit to the $V_{\rm rot} \sin i$
as there is clear evidence now for a macroturbulent broadening
mechanism in the atmospheres of hot supergiants
\citep{arXiv:0711.2264v1,2002MNRAS.336..577R}. The results from these
analyses would which would suggest a more likely rotational velocity
for Sher\,25 of 60~\kms. If we assume that the rotation axis is
aligned with the bi-polar lobes, then the inclination angle
$64^{\circ}$ \citep{1997ApJ...489L.153B} would suggest a $V_{\rm rot}
\simeq 70$~\kms, far lower than critical. However this does not rule
out the possibility of very fast rotation earlier in its lifetime. As
B-type supergiants are evolved from massive O-type main-sequence
stars, these progenitors often are rotating very rapidly and can be
near critical.

The overall stellar parameters of Sher\,25 (temperature, luminosity, 
mass-loss rate and photospheric abundances) are very much in line
with those derived for other Milky Way B-type supergiants and 
the star itself does not appear anomalous in any striking way 
\citep{2006A&A...446..279C,arXiv:astro-ph/0606717,arXiv:0711.1110}.
The nebular abundances presented here support this view point. 
Apart from its nebula, 
it appears as a typical B-type supergiant in a normal young 
star cluster. However as demonstrated clearly by Melena et al. (2008), 
Sher\,25 has an implied age of 4\,Myrs which is significantly older
than the most massive cluster members which are around 1\,Myr old. 
Sher\,25 also sits some 20$''$ from the core which contains
the bulk of the most massive stars. There is one other slightly 
evolved supergiant (Sher\,23), which also appears to be a similar
age to Sher\,25. So there is some evidence of an age spread in the
star-formation history of NGC\,3603 and the question of whether Sher\,25 was
formed in the core and ejected, or formed near were we see it now,
is still open.

\section{Stellar binarity analysis}\label{sec:bin}

\begin{figure}
    \centering 
    \epsfig{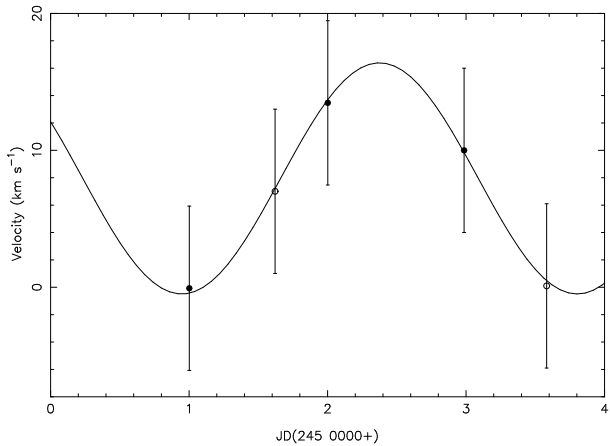} 
    \caption[Radial velocity shifts of Sher\,25 with the best
    $\chi^2$-fit for the second binary scenario.]{Radial velocity
    shifts of Sher\,25 over-plotted with the best $\chi^2$-fit sine
    function. The July and December velocity sets were shifted
    relative to each other arbitrarily in time and are shown with open
    and filled circles, respectively. The best fit has an amplitude of
    8.44~\kms, a systematic velocity of 7.95~\kms, a period of
    around 3\,d and a $\chi^2$ value of
    $1.10\times10^{-2}$.}\label{fig:sine2}
\end{figure}

The spectra listed in Table~\ref{tab:spec} were first corrected for
the heliocentric velocity and then cross-correlated with the
4547--4652\,\AA\ wavelength region of the spectrum of 1999 May 28 using
{\sc dipso}. The peak of the cross-correlation function was measured
by fitting one or more Gaussian profiles to it using the {\sc dipso}
package {\sc elf}. The resultant velocities are listed in
Table~\ref{tab:vcorr}. The July velocity set was shifted relative to
the December set arbitrarily in time so that they
coincided, revealing a sinusoidal velocity variation. The July and
December velocity sets are shown with open and filled circles in
Figure~\ref{fig:sine2}. In order to verify this variation a sine curve
was fitted to the data using a $\chi^2$-fitting algorithm. While
carrying out this $\chi^2$-fit the relative shift between the two
datasets was varied to find the best fit possible.  The July data set
was iteratively shifted in order to find the best $\chi^2$-fit. The
data was fitted by $7.95+8.44 \sin (2.20 t+2.63)$, which had
$\chi^2=1.10\times10^{-2}$, corresponding to a period of around 3\,d
and an amplitude of 8.44~\kms. The $\chi^2$ values are very low as a
result of the error of 6~\kms\ attached to each of the points, being
large and perhaps overestimated. The arbitrary amount that the July
data set was shifted by relative to the December dataset was
consistent with a full number of complete periods, within the
errors. The shifted velocities and best-fit sine function are plotted
in Figure~\ref{fig:sine2}.

\begin{table}
  \caption{Radial velocity of Sher\,25 with respect to the spectrum of 1999 May 28.}\label{tab:vcorr}
    \centering
    \begin{tabular}{lrrrrlll}
      \hline
      \hline\\[-11pt]
      Date & JD & Velocity with respect\\
      &  & to 1999 May 28\\
      & (245 0000+)&  (\kms)  \\[2pt]
      \hline\\[-11pt]
      1999 May 28 & 1327.00 &   0.00\\
      2004 Jul 05 & 3191.60 &   7.01\\
      2004 Jul 07 & 3193.56 &   0.10\\
      2004 Dec 30 & 3369.85 &  -0.07\\
      2004 Dec 31 & 3370.85 &  13.47\\
      2005 Jan 01 & 3371.83 &  10.00\\
      \hline
    \end{tabular}\\[5pt]
\end{table}

These radial velocity shifts could be indicative of an unseen binary
star. Using simple Keplerian arguments and the mass ($40\pm5$\msun) 
and radius ($60\pm15$\rsun ; from the models of Meynet et al. 2004)
for Sher\,25, we can put some
limits on the parameters of the system to ascertain if the
binary scenario is plausible. The parameters from the fit in
Figure~\ref{fig:sine2}, together with the lower limit of the radius
(45\,\rsun) as the lower limit of the separation, indicate that the
combined mass of the primary and secondary 
would have to be greater than 136\,\msun. 
This would suggest that the
secondary would be much more massive than Sher\,25, 
contrary to what we
observe. The short period, but low maximum velocity amplitude 
of Sher\,25 are hence inconsistent with a plausible close binary. 
Moving the centre of mass closer than the radius of Sher\,25 
might be attractive to slow merger hypotheses
\citep[e.g.,][]{2003fthp.conf...13P}. However, this is very 
speculative, as the dataset does not have enough temporal coverage to
say one way or another. An altogether different explanation for these
observations could be pulsations or some other natural variation in
the stellar envelope or, given that the velocity shifts are of the
same order as the uncertainties, an artifact of the wavelength
calibration. Note, however, the same velocity shift is not present in the
interstellar Ca\,{\sc ii} $\lambda$3933\,\AA\ line. Given this evidence, 
Sher\,25 certainly warrants further spectroscopic monitoring in the future.
  
\section{Sher\,25's variable wind}

Another interesting feature of Sher\,25 is evidence of a variable
wind. Figure~\ref{fig:wind} shows the H$\alpha$ and \SiIII\ profiles
of the Sher\,25 spectra, where the colours used are as follows: black =
1999 May 28, red = 2004 July~5, green = 2004 July~7,
blue~=~2004~December~30, light blue = 2004 December~31 and pink = 2005
January~1. The H$\alpha$ profiles, in all cases, show a variable
double peak, which is the nebular component of Sher\,25 superimposed
onto the stellar profile. The peaks of the profiles are probably
varying because of the different amounts of nebular emission that made
it into either the slit or the aperture, depending on their exact
position, or because of the sky subtraction. The UCLES (black line)
spectra are from a single-slit spectrograph, whereas FEROS is a
fibre-fed instrument. Paper~I noted that the
secondary peak bluewards of the rest wavelength of \ha\ in the UCLES
spectrum is a result of the imperfect nebular subtraction. FEROS has
two fibres, each of 2$''$ diameter, which are separated by a fixed
angular distance of 2.9$'$.  Because of field rotation, it is
highly probable that different amounts of the nebula were subtracted
from different FEROS observations, as a result of the sky fibre being
located at different positions in the extended and highly variable
background nebula.

The point of interest in this figure is the trough of the P-Cygni
profile, which varies significantly in time, indicating the
variability of the wind. The resolution of the spectra is good enough
to resolve the two components of the profile, therefore the variation
is considered to be real, as it will not be affected by the sky
subtraction. This variability is not seen in the profiles of the
photospheric \SiIII\ lines, which have been corrected for the observed
radial velocity offsets, indicating that it does indeed appear to
come from the wind.

\begin{figure}
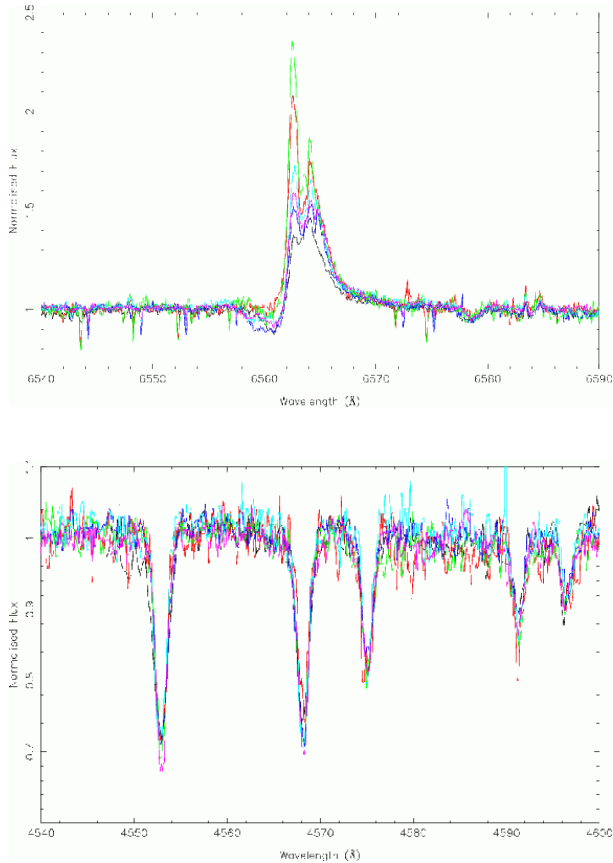

  \centering
  \begin{minipage}{0.5\textwidth}
    \centering
    \epsfig{file = Ha_b.epsi, width = 80mm}
  \end{minipage}\\[20pt]
  \begin{minipage}{0.5\textwidth}
    \centering
    \epsfig{file = SiIII_b.epsi, width = 80mm}
  \end{minipage}
  \caption[H$\alpha$ and \SiIII\ profiles of Sher\,25 showing a
  variable P-Cygni profile in H$\alpha$.]{H$\alpha$ and \SiIII\
  $\lambda\lambda$4553, 4568, 4575\,\AA\ profiles of Sher\,25 showing a
  variable P-Cygni profile in H$\alpha$, indicating a variable
  wind. No variations are seen in the \SiIII\ lines.  The
  colour sequence is: black = 1999 May 28, red = 2004
  July~5, green = 2004 July~7, blue~=~2004~December~30, light blue =
  2004 December 31 and pink = 2005 January~1.}
\end{figure}\label{fig:wind}

\section{Conclusions}\label{sec:conc}				    

Long-slit spectroscopy of the hourglass-shaped nebula surrounding
Sher\,25 was presented alongside that of the background nebula of
NGC\,3603. The nebular analysis confirmed the enhanced N
abundances suggested by \citet{1997ApJ...489L.153B}, but they are not
enhanced enough to indicate definitively that Sher\,25 has been through a RSG
phase. However, the N abundance was enhanced by a factor of 30 above
the background nebular values. A new analysis with the {\sc fastwind}
unified stellar atmosphere and wind code suggests that the
photospheric N and O abundances in Sher\,25 are consistent with the
nebular results. We therefore conclude that the nebula was ejected
from the star while it was a BSG. Probable variations in
the stellar radial velocity were discovered and it was shown that
there was a possible binary scenario for the variation with a period
and amplitude of 3\,d and 8.44~\kms. Using simple Keplerian arguments
it was shown that, if a companion exists, the combined mass of the
primary and the secondary would have to be greater than
136\,\msun. This is not a plausible scenario as no massive companion
is visible in the extensive imaging or spectroscopy of this object.
The masses might be more plausible if the separation of the binary is
less than the radius of Sher\,25, and we were witnessing a slow
merger. However, the uncertainties are comparable to the velocitiy
variations and additional observations are required to investigate
this aspect further.

\section*{Acknowledgments}
Based on observations at the European Southern Observatory at La
Silla, under programmes 071.D-0180 and 074.D-0021.  This work,
conducted as part of the award "Understanding the lives of massive
stars from birth to supernovae" (S.J. Smartt) made under the European
Heads of Research Councils and European Science Foundation EURYI
(European Young Investigator) Awards scheme, was supported by funds
from the Participating Organisations of EURYI and the EC Sixth
Framework Programme.  M. Hendry would like to thank PPARC and the
Leverhulme Trust for support. EDS is grateful for the hospitality of
the IoA during his sabbatical visit, and for partial support from the
University of Minnesota.  We thank Adam Burnley for obtaining some of
the FEROS observations presented, and Nathan Smith for a helpful
review of the paper and general discussions on blue supergiants.

\appendix
\section{Data tables of line fluxes and intensities}

\begin{table*}
\caption{Observed and corrected relative emission line fluxes and intensities, where the subscript denotes the sky subtraction applied.}\label{tab:fluxes}
\centering
\begin{tabular}{lrrrrrr}
\hline
\hline\\[-11pt]
& \multicolumn{3}{c}{EC$_{12}$}& \multicolumn{3}{c}{EC$_{123}$}\\
Ion & $F(\lambda)/F(\hb)$ & $I(\lambda)/I(\hb)$ & $W(\lambda)$ & $F(\lambda)/F(\hb)$ & $I(\lambda)/I(\hb)$ & $W(\lambda)$\\[2pt]
\hline\\[-10pt]
3727 \Oii	 &  0.109(0.015) & 0.725(0.099) &  25.50 &  0.049(0.015) &  0.346(0.105) &  11.06 \\
3868 \Neiii	 &  0.056(0.013) & 0.311(0.073) &  14.09 &  0.097(0.015) &  0.566(0.089) &  20.22 \\
3889 \Hei + H8   &  0.046(0.013) & 0.254(0.071) &  11.45 &  0.101(0.015) &  0.569(0.086) &  20.83 \\
3968 \Neiii + H7 &  0.041(0.011) & 0.199(0.051) &   9.48 &  0.042(0.013) &  0.205(0.065) &   8.11 \\
4026 \Hei	 &  0.004(0.011) & 0.016(0.047) &   0.80 &  0.006(0.013) &  0.026(0.059) &   1.05 \\
4101 \hd	 &  0.073(0.008) & 0.287(0.033) &  13.13 &  0.073(0.010) &  0.288(0.042) &  13.01 \\
4340 \hg	 &  0.227(0.010) & 0.575(0.026) &  33.82 &  0.222(0.012) &  0.572(0.032) &  32.96 \\
4471 \Hei	 &  0.023(0.016) & 0.045(0.031) &   3.48 &  0.022(0.018) &  0.045(0.035) &   2.79 \\
4861 \hb	 &  1.000(0.023) & 1.000(0.016) &  97.70 &  1.000(0.026) &  1.000(0.018) &  94.23 \\
4921 \Hei	 &  0.007(0.007) & 0.006(0.007) &   0.78 &  0.010(0.011) &  0.009(0.010) &   0.91 \\
4959 \Oiii	 &  0.872(0.021) & 0.748(0.014) &  78.05 &  0.812(0.023) &  0.696(0.015) &  70.65 \\
5007 \Oiii	 &  2.829(0.056) & 2.259(0.026) & 242.91 &  2.648(0.057) &  2.110(0.027) & 221.84 \\
5755 \Nii	 &  0.060(0.017) & 0.020(0.006) &   3.34 &  0.066(0.017) &  0.021(0.006) &   3.43 \\
5876 \Hei	 &  0.628(0.019) & 0.190(0.006) &  34.44 &  0.677(0.023) &  0.198(0.007) &  34.95 \\
6548 \Nii	 &  5.925(0.114) & 1.041(0.033) & 262.28 &  7.007(0.147) &  1.176(0.045) & 288.81 \\
6563 \ha	 & 20.235(0.387) & 3.515(0.110) & 891.87 & 21.349(0.445) &  3.539(0.136) & 876.02 \\
6584 \Nii	 & 17.775(0.340) & 3.037(0.096) & 777.98 & 21.018(0.438) &  3.427(0.133) & 855.75 \\
6678 \Hei	 &  0.347(0.015) & 0.055(0.003) &  14.71 &  0.376(0.018) &  0.057(0.003) &  14.78 \\
6717 \Sii	 &  0.441(0.016) & 0.068(0.003) &  18.41 &  0.399(0.018) &  0.059(0.003) &  15.46 \\
6731 \Sii	 &  0.522(0.017) & 0.080(0.003) &  21.71 &  0.527(0.020) &  0.076(0.004) &  20.32 \\
7065 \Hei	 &  0.378(0.025) & 0.044(0.003) &  14.21 &  0.415(0.030) &  0.046(0.004) &  14.40 \\
7136 \Ariii      &  1.145(0.033) & 0.127(0.006) &  43.04 &  1.211(0.039) &  0.127(0.007) &  41.75 \\
$F(\hb)\,(\rm{erg}\,cm^{-2}\,s^{-1})$ & \multicolumn{3}{c}{$1.45\times10^{-14}$} & \multicolumn{3}{c}{$1.208\times10^{-14}$}\\
$C(\hb)$ & \multicolumn{3}{c}{2.54(0.04)} & \multicolumn{3}{c}{2.61(0.05)}\\   
$W_{\rm abs}\,($\AA$)$ & \multicolumn{3}{c}{0.16(2.09)} & \multicolumn{3}{c}{-0.22(2.81)}\\
\hline	         					         
\end{tabular}\\[5pt]
\normalsize 		       
\end{table*}

\begin{table*}
\centering
\begin{tabular}{lrrrrrr}
\multicolumn{7}{c}{{\tablename} \thetable{} -- Continued} \\[4pt]
\hline
\hline\\[-11pt]
& \multicolumn{3}{c}{WC$_{12}$}&\multicolumn{3}{c}{WC$_{123}$}\\
Ion & $F(\lambda)/F(\hb)$ & $I(\lambda)/I(\hb)$ & $W(\lambda)$ & $F(\lambda)/F(\hb)$ & $I(\lambda)/I(\hb)$ & $W(\lambda)$\\[2pt]
\hline\\[-10pt]
3727 \Oii	 &  0.220(0.013) &  0.822(0.054)  &   2.83 &  0.226(0.015) & 0.836(0.059) &  2.78 \\
4340 \hg	 &  0.065(0.005) &  0.512(0.012)  &   0.56 &  0.056(0.005) & 0.512(0.012) &  0.46 \\
4861 \hb	 &  1.000(0.036) &  1.000(0.019)  &   6.98 &  1.000(0.037) & 1.000(0.020) &  6.62 \\
4959 \Oiii	 &  0.648(0.032) &  0.440(0.019)  &   4.36 &  0.611(0.034) & 0.409(0.020) &  3.89 \\
5007 \Oiii	 &  1.901(0.058) &  1.216(0.021)  &  12.42 &  1.777(0.057) & 1.123(0.021) & 10.99 \\
5755 \Nii	 &  0.077(0.032) &  0.024(0.010)  &   0.35 &  0.081(0.033) & 0.025(0.010) &  0.35 \\
5876 \Hei	 &  0.523(0.029) &  0.149(0.008)  &   2.34 &  0.541(0.032) & 0.152(0.009) &  2.29 \\
6548 \Nii	 &  8.089(0.226) &  1.469(0.050)  &  33.43 &  8.530(0.247) & 1.531(0.053) & 33.17 \\
6563 \ha	 & 18.563(0.505) &  3.428(0.116)  &  76.41 & 18.750(0.531) & 3.428(0.118) & 72.68 \\
6584 \Nii	 & 24.276(0.659) &  4.306(0.147)  &  99.38 & 25.597(0.722) & 4.488(0.156) & 98.76 \\
6678 \Hei	 &  0.220(0.060) &  0.037(0.010)  &   0.88 &  0.212(0.062) & 0.035(0.010) &  0.80 \\
6717 \Sii	 &  0.406(0.062) &  0.066(0.010)  &   1.61 &  0.393(0.066) & 0.063(0.011) &  1.47 \\
6731 \Sii	 &  0.576(0.063) &  0.093(0.010)  &   2.27 &  0.587(0.067) & 0.094(0.011) &  2.19 \\
7065 \Hei	 &  0.229(0.051) &  0.030(0.007)  &   0.85 &  0.225(0.052) & 0.029(0.007) &  0.79 \\
7136 \Ariii      &  0.682(0.057) &  0.085(0.008)  &   2.51 &  0.653(0.059) & 0.080(0.008) &  2.27 \\
$F(\hb) (\rm{erg}\,cm^{-2}\,s^{-1})$ & \multicolumn{3}{c}{$5.281\times10^{-14}$} & \multicolumn{3}{c}{$4.985\times10^{-14}$}\\
$C(\hb)$ & \multicolumn{3}{c}{2.11(0.05)} & \multicolumn{3}{c}{2.11(0.05)}\\
$W_{\rm abs}\,($\AA$)$ & \multicolumn{3}{c}{2.09(0.14)} & \multicolumn{3}{c}{2.08(0.14)}\\
\hline
\end{tabular}\\[5pt]
\scriptsize EC = East cloud, WC = West cloud, R1 = Ring 1, R2 = Ring 2, NA = Nebula A, NB = Nebula B, NC = Nebula C and NX = Nebula X
\normalsize  		       
\end{table*}

\begin{table*}
\centering
\begin{tabular}{lrrrrrr}
\multicolumn{7}{c}{{\tablename} \thetable{} -- Continued} \\[4pt]
\hline
\hline\\[-11pt]
& \multicolumn{3}{c}{R1}& \multicolumn{3}{c}{R1+R2}\\
Ion & $F(\lambda)/F(\hb)$ & $I(\lambda)/I(\hb)$ & $W(\lambda)$ & $F(\lambda)/F(\hb)$ & $I(\lambda)/I(\hb)$ & $W(\lambda)$\\[2pt]
\hline\\[-10pt]
3727 \Oii	 &  0.110(0.023)  & 0.485(0.101) &   6.92 &  0.047(0.020) & 0.227(0.095) &   3.05 \\
3868 \Neiii	 &  0.045(0.016)  & 0.172(0.060) &   2.77 &  0.054(0.017) & 0.222(0.072) &   3.38 \\
3889 \Hei + H8   &  0.061(0.016)  & 0.368(0.059) &   3.83 &  0.064(0.017) & 0.398(0.072) &   4.04 \\
3968 \Neiii + H7 &  0.050(0.013)  & 0.289(0.046) &   3.29 &  0.040(0.014) & 0.270(0.052) &   2.55 \\
4101 \hd	 &  0.070(0.007)  & 0.302(0.021) &   4.45 &  0.064(0.015) & 0.303(0.046) &   4.03 \\
4340 \hg	 &  0.224(0.016)  & 0.525(0.033) &  14.35 &  0.212(0.018) & 0.521(0.038) &  13.03 \\
4471 \Hei	 &  0.035(0.021)  & 0.058(0.034) &   2.08 &  --           & --           &   --   \\
4861 \hb	 &  1.000(0.025)  & 1.000(0.017) &  51.73 &  1.000(0.029) & 1.000(0.019) &  47.70 \\
4959 \Oiii	 &  0.592(0.019)  & 0.501(0.014) &  29.03 &  0.647(0.023) & 0.543(0.016) &  28.89 \\
5007 \Oiii	 &  1.843(0.041)  & 1.474(0.020) &  88.04 &  2.027(0.049) & 1.601(0.023) &  87.75 \\
5538 \Cliii	 &  0.011(0.013)  & 0.005(0.006) &   0.41 &  0.010(0.013) & 0.005(0.006) &   0.33 \\
5876 \Hei	 &  0.590(0.020)  & 0.216(0.008) &  20.74 &  0.623(0.026) & 0.214(0.010) &  19.15 \\
6548 \Nii	 &  2.953(0.064)  & 0.698(0.027) &  98.60 &  3.089(0.073) & 0.669(0.030) &  86.86 \\
6563 \ha	 & 14.465(0.297)  & 3.403(0.131) & 481.54 & 15.926(0.360) & 3.431(0.152) & 445.99 \\
6584 \Nii	 &  8.859(0.183)  & 2.047(0.080) & 293.91 &  9.266(0.210) & 1.959(0.088) & 258.23 \\
6678 \Hei	 &  0.284(0.021)  & 0.062(0.005) &   9.26 &  0.307(0.022) & 0.061(0.005) &   8.37 \\
6717 \Sii	 &  0.269(0.024)  & 0.057(0.005) &   8.70 &  0.264(0.026) & 0.051(0.005) &   7.13 \\
6731 \Sii	 &  0.287(0.024)  & 0.061(0.005) &   9.28 &  0.290(0.026) & 0.056(0.005) &   7.79 \\
7065 \Hei	 &  0.306(0.024)  & 0.052(0.005) &   9.48 &  0.347(0.022) & 0.053(0.004) &   8.73 \\
7136 \Ariii      &  0.609(0.028)  & 0.100(0.006) &  19.15 &  0.671(0.026) & 0.098(0.006) &  16.42 \\
$F(\hb) (\rm{erg}\,cm^{-2}\,s^{-1})$ & \multicolumn{3}{c}{$1.046\times10^{-14}$} & \multicolumn{3}{c}{$1.431\times10^{-14}$}\\
$C(\hb)$ & \multicolumn{3}{c}{2.04(0.05)} & \multicolumn{3}{c}{2.17(0.06)}\\
$W_{\rm abs}\,($\AA$)$ & \multicolumn{3}{c}{2.33(0.59)} & \multicolumn{3}{c}{2.20(1.20)}\\
\hline
\end{tabular}\\[5pt]
\scriptsize EC = East cloud, WC = West cloud, R1 = Ring 1, R2 = Ring 2, NA = Nebula A, NB = Nebula B, NC = Nebula C and NX = Nebula X
\normalsize 	     
\end{table*}
  		     
\begin{table*}
\centering
\begin{tabular}{lrrrrrr}
\multicolumn{7}{c}{{\tablename} \thetable{} -- Continued} \\[4pt]
\hline
\hline\\[-11pt]
& \multicolumn{3}{c}{NA$_{12}$}&\multicolumn{3}{c}{NA$_{123}$}\\
Ion & $F(\lambda)/F(\hb)$ & $I(\lambda)/I(\hb)$ & $W(\lambda)$ & $F(\lambda)/F(\hb)$ & $I(\lambda)/I(\hb)$ & $W(\lambda)$\\[2pt]
\hline\\[-10pt]
3727 \Oii	 &  0.096(0.002) & 0.398(0.011) &   16.42 &  0.090(0.003) & 0.378(0.013) &   15.40 \\
3798 H10	 &  0.003(0.003) & 0.045(0.011) &    0.54 &  0.004(0.003) & 0.041(0.013) &    0.63 \\
3820 \Hei	 &  0.006(0.003) & 0.024(0.010) &    0.99 &  0.007(0.003) & 0.027(0.012) &    1.08 \\
3835 H9	         &  0.012(0.003) & 0.084(0.010) &    1.83 &  0.013(0.003) & 0.080(0.012) &    2.00 \\ 
3868 \Neiii	 &  0.140(0.003) & 0.508(0.013) &   20.67 &  0.144(0.004) & 0.525(0.015) &   21.50 \\
3889 \Hei + H8   &  0.051(0.003) & 0.228(0.010) &    7.36 &  0.052(0.003) & 0.225(0.012) &    7.74 \\
3968 \Neiii + H7 &  0.073(0.003) & 0.284(0.009) &   10.22 &  0.077(0.003) & 0.287(0.011) &   10.85 \\
4026 \Hei	 &  0.005(0.003) & 0.015(0.008) &    0.69 &  0.006(0.003) & 0.017(0.009) &    0.79 \\
4101 \hd	 &  0.091(0.002) & 0.289(0.007) &   12.44 &  0.091(0.003) & 0.284(0.009) &   12.46 \\
4340 \hg	 &  0.274(0.004) & 0.572(0.007) &   38.82 &  0.274(0.005) & 0.570(0.009) &   38.83 \\
4363 \Oiii	 &  0.009(0.002) & 0.016(0.003) &    1.21 &  0.009(0.003) & 0.017(0.005) &    1.23 \\
4387 \Hei	 &  0.002(0.003) & 0.003(0.005) &    0.22 &  0.002(0.003) & 0.005(0.006) &    0.35 \\
4471 \Hei	 &  0.034(0.003) & 0.056(0.005) &    4.75 &  0.034(0.003) & 0.056(0.006) &    4.70 \\
4686 \Heii	 &  0.014(0.004) & 0.017(0.005) &    1.90 &  0.014(0.005) & 0.018(0.006) &    2.02 \\
4713 \Ariv + \Hei &  0.024(0.004) & 0.028(0.005) &    3.30 &  0.024(0.005) & 0.029(0.006) &    3.39 \\
4740 \Ariv	 &  0.008(0.004) & 0.009(0.005) &    1.09 &  0.008(0.005) & 0.010(0.005) &    1.16 \\
4861 \hb	 &  1.000(0.015) & 1.000(0.010) &  135.77 &  1.000(0.015) & 1.000(0.010) &  136.42 \\
4921 \Hei	 &  0.012(0.003) & 0.011(0.003) &    1.65 &  0.013(0.003) & 0.012(0.003) &    1.70 \\
4959 \Oiii	 &  1.908(0.028) & 1.676(0.017) &  251.16 &  1.920(0.028) & 1.691(0.017) &  254.34 \\
5007 \Oiii	 &  6.038(0.088) & 5.024(0.051) &  783.03 &  6.076(0.088) & 5.069(0.052) &  793.41 \\
5518 \Cliii	 &  0.010(0.003) & 0.005(0.001) &    1.11 &  0.010(0.003) & 0.005(0.002) &    1.11 \\
5538 \Cliii	 &  0.008(0.003) & 0.004(0.001) &    0.90 &  0.008(0.003) & 0.004(0.002) &    0.88 \\
5876 \Hei	 &  0.407(0.006) & 0.162(0.002) &   45.30 &  0.407(0.006) & 0.162(0.002) &   45.67 \\
6312 \Siii	 &  0.038(0.003) & 0.011(0.001) &    4.15 &  0.037(0.004) & 0.011(0.001) &    4.05 \\
6563 \ha	 & 12.948(0.188) & 3.395(0.057) & 1358.42 & 12.902(0.187) & 3.373(0.060) & 1363.58 \\
6584 \Nii	 &  0.272(0.005) & 0.070(0.001) &   28.41 &  0.247(0.004) & 0.064(0.001) &   25.97 \\
6678 \Hei	 &  0.192(0.004) & 0.047(0.001) &   19.74 &  0.191(0.004) & 0.047(0.001) &   19.76 \\
6717 \Sii	 &  0.056(0.003) & 0.013(0.001) &    5.70 &  0.045(0.003) & 0.011(0.001) &    4.62 \\
6731 \Sii	 &  0.048(0.003) & 0.011(0.001) &    4.86 &  0.039(0.003) & 0.009(0.001) &    3.99 \\
7065 \Hei	 &  0.281(0.007) & 0.055(0.002) &   27.26 &  0.284(0.010) & 0.055(0.002) &   27.93 \\
7136 \Ariii      &  0.910(0.015) & 0.170(0.004) &   86.96 &  0.921(0.016) & 0.171(0.004) &   91.13 \\
$F(\hb) (\rm{erg}\,cm^{-2}\,s^{-1})$ & \multicolumn{3}{c}{$8.225\times10^{-14}$} & \multicolumn{3}{c}{$8.084\times10^{-14}$}\\
$C(\hb)$ & \multicolumn{3}{c}{1.92(0.02)} & \multicolumn{3}{c}{1.93(0.02)}\\
$W_{\rm abs}\,($\AA$)$ & \multicolumn{3}{c}{1.99(0.43)} & \multicolumn{3}{c}{1.56(0.52)}\\
\hline
\end{tabular}\\[5pt]
\scriptsize EC = East cloud, WC = West cloud, R1 = Ring 1, R2 = Ring 2, NA = Nebula A, NB = Nebula B, NC = Nebula C and NX = Nebula X
\normalsize 	     
\end{table*}

\begin{table*}
\centering
\begin{tabular}{lrrrrrr}
\multicolumn{7}{c}{{\tablename} \thetable{} -- Continued} \\[4pt]
\hline
\hline\\[-11pt]
& \multicolumn{3}{c}{NB$_{12}$}&\multicolumn{3}{c}{NB$_{123}$}\\
Ion & $F(\lambda)/F(\hb)$ & $I(\lambda)/I(\hb)$ & $W(\lambda)$ & $F(\lambda)/F(\hb)$ & $I(\lambda)/I(\hb)$ & $W(\lambda)$\\[2pt]
\hline\\[-10pt]
3727 \Oii	 &  0.176(0.003)  & 0.860(0.019) &   46.61 &  0.173(0.003) & 0.832(0.019) &   45.96 \\
3798 H10	 &  0.005(0.004)  & 0.036(0.016) &    1.26 &  0.006(0.004) & 0.046(0.016) &    1.52 \\
3820 \Hei	 &  0.003(0.004)  & 0.015(0.016) &    0.79 &  0.004(0.004) & 0.016(0.016) &    0.86 \\
3835 H9	         &  0.012(0.004)  & 0.069(0.015) &    2.79 &  0.012(0.004) & 0.074(0.015) &    2.81 \\ 
3868 \Neiii	 &  0.131(0.004)  & 0.550(0.017) &   28.85 &  0.131(0.004) & 0.543(0.017) &   29.02 \\
3889 \Hei + H8   &  0.045(0.004)  & 0.203(0.015) &    9.67 &  0.044(0.004) & 0.205(0.015) &    9.53 \\
3968 \Neiii + H7 &  0.078(0.004)  & 0.309(0.013) &   15.97 &  0.078(0.004) & 0.312(0.013) &   15.92 \\
4026 \Hei	 &  0.007(0.003)  & 0.025(0.011) &    1.45 &  0.007(0.003) & 0.023(0.011) &    1.38 \\
4101 \hd	 &  0.092(0.002)  & 0.300(0.006) &   18.62 &  0.090(0.002) & 0.298(0.006) &   18.08 \\
4340 \hg	 &  0.260(0.004)  & 0.572(0.008) &   54.19 &  0.259(0.004) & 0.569(0.008) &   53.65 \\
4363 \Oiii	 &  0.009(0.002)  & 0.018(0.004) &    1.84 &  0.008(0.002) & 0.016(0.004) &    1.60 \\
4387 \Hei	 &  0.003(0.004)  & 0.006(0.007) &    0.61 &  0.002(0.004) & 0.004(0.007) &    0.42 \\
4471 \Hei	 &  0.032(0.004)  & 0.056(0.006) &    6.52 &  0.032(0.004) & 0.056(0.006) &    6.57 \\
4686 \Heii	 &  0.010(0.005)  & 0.013(0.007) &    2.10 &  0.010(0.005) & 0.013(0.007) &    2.14 \\
4713 \Ariv + \Hei &  0.016(0.005)  & 0.020(0.007) &    3.25 &  0.016(0.005) & 0.019(0.007) &    3.23 \\
4740 \Ariv	 &  0.004(0.005)  & 0.005(0.006) &    0.88 &  0.004(0.005) & 0.005(0.006) &    0.89 \\
4861 \hb	 &  1.000(0.015)  & 1.000(0.010) &  195.49 &  1.000(0.015) & 1.000(0.010) &  197.02 \\
4921 \Hei	 &  0.013(0.004)  & 0.012(0.003) &    2.44 &  0.013(0.004) & 0.012(0.004) &    2.60 \\
4959 \Oiii	 &  1.965(0.028)  & 1.721(0.017) &  368.75 &  1.977(0.029) & 1.730(0.018) &  377.13 \\
5007 \Oiii	 &  6.271(0.090)  & 5.174(0.053) & 1153.32 &  6.306(0.091) & 5.200(0.053) & 1183.96 \\
5518 \Cliii	 &  0.011(0.004)  & 0.005(0.002) &    1.76 &  0.012(0.004) & 0.006(0.002) &    1.86 \\
5538 \Cliii	 &  0.008(0.004)  & 0.004(0.002) &    1.26 &  0.008(0.004) & 0.004(0.002) &    1.31 \\
5755 \Nii	 &  0.008(0.002)  & 0.003(0.001) &    1.15 &  0.007(0.002) & 0.003(0.001) &    1.14 \\
5876 \Hei	 &  0.426(0.006)  & 0.156(0.002) &   63.24 &  0.427(0.006) & 0.157(0.002) &   63.98 \\
6312 \Siii	 &  0.059(0.004)  & 0.016(0.001) &    8.16 &  0.058(0.004) & 0.016(0.001) &    8.03 \\
6563 \ha	 & 14.882(0.214)  & 3.419(0.058) & 1897.09 & 14.715(0.212) & 3.421(0.058) & 1888.67 \\
6584 \Nii	 &  0.711(0.011)  & 0.161(0.003) &   90.28 &  0.954(0.015) & 0.219(0.004) &  121.75 \\
6678 \Hei	 &  0.208(0.006)  & 0.044(0.001) &   25.75 &  0.207(0.006) & 0.045(0.001) &   25.74 \\
6717 \Sii	 &  0.105(0.006)  & 0.022(0.001) &   12.87 &  0.098(0.006) & 0.021(0.001) &   12.00 \\
6731 \Sii	 &  0.151(0.006)  & 0.031(0.001) &   18.41 &  0.145(0.006) & 0.030(0.001) &   17.81 \\
7065 \Hei	 &  0.319(0.010)  & 0.053(0.002) &   35.51 &  0.321(0.010) & 0.054(0.002) &   36.05 \\
7136 \Ariii      &  1.143(0.019)  & 0.180(0.004) &  127.06 &  1.146(0.019) & 0.184(0.004) &  128.06 \\
$F(\hb) (\rm{erg}\,cm^{-2}\,s^{-1})$ & \multicolumn{3}{c}{$1.100\times10^{-13}$} & \multicolumn{3}{c}{$1.086\times10^{-13}$}\\
$C(\hb)$ & \multicolumn{3}{c}{2.13(0.02)} & \multicolumn{3}{c}{2.11(0.02)}\\
$W_{\rm abs}\,($\AA$)$ & \multicolumn{3}{c}{1.03(0.57)} & \multicolumn{3}{c}{1.49(0.57)}\\
\hline
\end{tabular}\\[5pt]
\scriptsize EC = East cloud, WC = West cloud, R1 = Ring 1, R2 = Ring 2, NA = Nebula A, NB = Nebula B, NC = Nebula C and NX = Nebula X
\normalsize 	     
\end{table*}
 		     
\begin{table*}
\centering
\begin{tabular}{lrrrrrr}
\multicolumn{7}{c}{{\tablename} \thetable{} -- Continued} \\[4pt]
\hline
\hline\\[-11pt]
& \multicolumn{3}{c}{NC$_{12}$}&\multicolumn{3}{c}{NC$_{123}$}\\
Ion & $F(\lambda)/F(\hb)$ & $I(\lambda)/I(\hb)$ & $W(\lambda)$ & $F(\lambda)/F(\hb)$ & $I(\lambda)/I(\hb)$ & $W(\lambda)$\\[2pt]
\hline\\[-10pt]
3727 \Oii	 &  0.189(0.005)  & 0.997(0.029) &   75.86 &  0.183(0.005) & 0.983(0.030) &   77.01 \\
3798 H10	 &  0.012(0.003)  & 0.018(0.015) &    5.31 &  0.013(0.003) & 0.003(0.016) &    5.61 \\
3820 \Hei	 &  0.002(0.003)  & 0.008(0.014) &    0.69 &  0.001(0.003) & 0.003(0.015) &    0.24 \\
3835 H9	         &  0.023(0.003)  & 0.060(0.014) &   10.02 &  0.022(0.003) & 0.036(0.015) &    9.99 \\ 
3868 \Neiii	 &  0.115(0.003)  & 0.518(0.017) &   48.79 &  0.115(0.004) & 0.527(0.017) &   50.91 \\
3889 \Hei + H8   &  0.066(0.003)  & 0.231(0.014) &   27.88 &  0.064(0.003) & 0.204(0.015) &   28.45 \\
3968 \Neiii + H7 &  0.089(0.003)  & 0.300(0.013) &   36.75 &  0.091(0.003) & 0.294(0.014) &   40.08 \\
4026 \Hei	 &  0.008(0.003)  & 0.029(0.011) &    3.25 &  0.009(0.003) & 0.033(0.012) &    3.84 \\
4101 \hd	 &  0.103(0.003)  & 0.292(0.008) &   39.96 &  0.106(0.003) & 0.290(0.009) &   44.35 \\
4340 \hg	 &  0.270(0.004)  & 0.576(0.009) &   98.55 &  0.270(0.004) & 0.573(0.009) &  104.99 \\
4363 \Oiii	 &  0.011(0.002)  & 0.023(0.004) &    3.84 &  0.010(0.002) & 0.022(0.004) &    3.89 \\
4387 \Hei	 &  0.003(0.003)  & 0.006(0.006) &    1.04 &  0.003(0.003) & 0.007(0.007) &    1.23 \\
4471 \Hei	 &  0.030(0.003)  & 0.055(0.006) &   10.24 &  0.030(0.003) & 0.055(0.006) &   10.83 \\
4686 \Heii	 &  0.008(0.004)  & 0.010(0.006) &    2.47 &  0.008(0.004) & 0.011(0.006) &    2.68 \\
4713 \Ariv + \Hei &  0.012(0.004)  & 0.016(0.006) &    3.86 &  0.012(0.004) & 0.015(0.005) &    3.85 \\
4740 \Ariv	 &  0.004(0.004)  & 0.005(0.005) &    1.21 &  0.004(0.004) & 0.005(0.005) &    1.29 \\
4861 \hb	 &  1.000(0.014)  & 1.000(0.010) &  279.32 &  1.000(0.014) & 1.000(0.011) &  296.67 \\
4921 \Hei	 &  0.016(0.003)  & 0.015(0.003) &    4.30 &  0.017(0.004) & 0.016(0.003) &    4.79 \\
4959 \Oiii	 &  1.864(0.027)  & 1.668(0.017) &  491.44 &  1.892(0.027) & 1.705(0.017) &  534.06 \\
5007 \Oiii	 &  5.960(0.085)  & 5.016(0.051) & 1527.73 &  6.046(0.087) & 5.120(0.052) & 1667.31 \\
5518 \Cliii	 &  0.011(0.003)  & 0.005(0.002) &    2.27 &  0.012(0.004) & 0.006(0.002) &    2.69 \\
5538 \Cliii	 &  0.008(0.003)  & 0.004(0.002) &    1.65 &  0.009(0.004) & 0.004(0.002) &    1.85 \\
5755 \Nii	 &  0.005(0.003)  & 0.002(0.001) &    0.96 &  0.004(0.003) & 0.002(0.001) &    0.84 \\
5876 \Hei	 &  0.433(0.007)  & 0.157(0.002) &   78.77 &  0.433(0.007) & 0.157(0.003) &   82.68 \\
6312 \Siii	 &  0.055(0.004)  & 0.015(0.001) &    8.81 &  0.048(0.004) & 0.013(0.001) &    7.91 \\
6563 \ha	 & 15.193(0.217)  & 3.410(0.063) & 2176.27 & 15.255(0.218) & 3.407(0.065) & 2283.83 \\
6584 \Nii	 &  0.624(0.009)  & 0.139(0.003) &   88.88 &  0.503(0.008) & 0.111(0.002) &   74.69 \\
6678 \Hei	 &  0.216(0.004)  & 0.045(0.001) &   29.81 &  0.216(0.004) & 0.045(0.001) &   31.02 \\
6717 \Sii	 &  0.116(0.004)  & 0.023(0.001) &   15.75 &  0.094(0.004) & 0.019(0.001) &   13.30 \\
6731 \Sii	 &  0.107(0.003)  & 0.022(0.001) &   14.51 &  0.091(0.004) & 0.018(0.001) &   12.74 \\
7065 \Hei	 &  0.275(0.009)  & 0.044(0.002) &   34.01 &  0.279(0.010) & 0.044(0.002) &   35.83 \\
7136 \Ariii      &  1.135(0.018)  & 0.174(0.004) &  139.37 &  1.147(0.019) & 0.174(0.004) &  145.56 \\
$F(\hb) (\rm{erg}\,cm^{-2}\,s^{-1})$ & \multicolumn{3}{c}{$2.958\times10^{-13}$} & \multicolumn{3}{c}{$0.2843\times10^{-12}$}\\
$C(\hb)$ & \multicolumn{3}{c}{2.19(0.02)} & \multicolumn{3}{c}{2.21(0.02)}\\
$W_{\rm abs}\,($\AA$)$ & \multicolumn{3}{c}{-5.65(1.47)} & \multicolumn{3}{c}{-8.26(1.76)}\\
\hline
\end{tabular}\\[5pt]
\scriptsize EC = East cloud, WC = West cloud, R1 = Ring 1, R2 = Ring 2, NA = Nebula A, NB = Nebula B, NC = Nebula C and NX = Nebula X
\normalsize 	     
\end{table*}
\begin{table*}
\begin{minipage}{.55\textwidth}
\centering
\begin{tabular}{lrrr}
\multicolumn{4}{c}{{\tablename} \thetable{} -- Continued} \\[4pt]
\hline
\hline\\[-11pt]
& \multicolumn{3}{c}{NX}\\
Ion & $F(\lambda)/F(\hb)$ & $I(\lambda)/I(\hb)$ & $W(\lambda)$\\[2pt]
\hline\\[-10pt]
3727 \Oii	 &  0.252(0.016)  & 1.504(0.103) &  226.84 \\
3835 H9	         &  0.020(0.009)  & 0.088(0.048) &   24.76 \\ 
3868 \Neiii	 &  0.070(0.009)  & 0.351(0.047) &   96.10 \\
3889 \Hei + H8   &  0.052(0.009)  & 0.236(0.045) &   75.71 \\
3968 \Neiii + H7 &  0.077(0.009)  & 0.323(0.041) &  119.99 \\
4026 \Hei	 &  0.011(0.009)  & 0.043(0.037) &   14.64 \\
4101 \hd	 &  0.090(0.008)  & 0.304(0.029) &   86.20 \\
4340 \hg	 &  0.235(0.010)  & 0.550(0.025) &  204.70 \\
4363 \Oiii	 &  0.006(0.010)  & --           &    --   \\
4387 \Hei	 &  0.005(0.010)  & 0.011(0.021) &    4.30 \\
4471 \Hei	 &  0.029(0.011)  & 0.055(0.021) &   22.24 \\
4861 \hb	 &  1.000(0.018)  & 1.000(0.013) &  533.12 \\
4921 \Hei	 &  0.011(0.010)  & 0.010(0.009) &    5.48 \\
4959 \Oiii	 &  1.543(0.026)  & 1.351(0.016) &  730.27 \\
5007 \Oiii	 &  4.951(0.081)  & 4.055(0.044) & 2219.13 \\
5518 \Cliii	 &  0.011(0.009)  & 0.005(0.004) &    3.82 \\
5538 \Cliii	 &  0.013(0.009)  & 0.006(0.004) &    4.00 \\
5876 \Hei	 &  0.468(0.013)  & 0.155(0.005) &  133.25 \\
6312 \Siii	 &  0.065(0.011)  & 0.015(0.003) &   13.52 \\
6563 \ha	 & 17.360(0.284)  & 3.410(0.105) & 3236.00 \\
6584 \Nii	 &  1.238(0.022)  & 0.240(0.008) &  228.98 \\
6678 \Hei	 &  0.245(0.011)  & 0.044(0.002) &   43.75 \\
6717 \Sii	 &  0.294(0.012)  & 0.052(0.003) &   51.73 \\
6731 \Sii	 &  0.239(0.012)  & 0.042(0.002) &   41.80 \\
7065 \Hei	 &  0.303(0.015)  & 0.041(0.002) &   48.62 \\
7136 \Ariii      &  1.234(0.025)  & 0.160(0.006) &  191.13 \\
$F(\hb) (\rm{erg}\,cm^{-2}\,s^{-1})$ &  \multicolumn{3}{c}{$1.481\times10^{-13}$}\\
$C(\hb)$ & \multicolumn{3}{c}{2.37(0.04)}\\
$W_{\rm abs}\,($\AA$)$ &  \multicolumn{3}{c}{-4.80(9.26)}\\
\hline         					         
\end{tabular}\\[5pt]
\scriptsize EC = East cloud, WC = West cloud, R1 = Ring 1, R2 = Ring 2, NA = Nebula A, NB = Nebula B, NC = Nebula C and NX = Nebula X	
\normalsize 	     
\end{minipage}
\end{table*}

\section{Tables of ionic and total nebular abundances}

\begin{table*}
\caption{Ionic and total abundances for the Sher\,25 nebula.}\label{tab:Sher25}
\centering
\begin{tabular}{lrrrrrr}
\hline
\hline\\[-11pt]
Property & EC$_{12}$ & EC$_{123}$ & WC$_{12}$ & WC$_{123}$ & R1 & R1+R2\\
\hline\\[-10pt]
  $T$(\ion{O}{iii})\,(K) (measured)          &  --            &  --             &  --           &  --           &  --         &  --  \\
  $T$(\ion{N}{ii})\,(K) (measured)         &   7900(700)    &    7700(700)    &   7500(100)   &   7500(1000)  & --          & -- \\[10pt]

  $T$(\ion{O}{iii})\,(K)         &   7000(600)    &    6800(600)    &   6400(900)   &   6400(900)  & 6700(2000)   & 6700(2000)	 \\
  $T$(\ion{O}{ii})\,(K)         &   7900(700)    &    7700(700)    &   7500(1000)   &   7500(1000) & 7700(2300)  & 7700(2300)	 \\
  $T$(\ion{S}{iii})\,(K)         &   7500(700)    &    7300(600)    &   7000(1000)   &   7000(900)  & 7200(2200)  & 7200(2200)	 \\
  O$^+$/H$^{+}$ ($10^5$)             &   9.07(4.48)   &    4.81(2.64)   &  13.94(10.59)  &  14.10(10.40) &   7.22(11.87) &   3.38(5.69) \\
  O$^{++}$/H$^{+}$ ($10^5$)          &  35.45(14.33)  &   38.20(15.01)  &  29.81(19.82)  &  27.38(17.66) &  29.59(41.84) &  32.12(45.41) \\
  O/H ($10^4$)               &   4.45(1.50)   &    4.30(1.52)   &   4.38(2.25)  &   4.15(2.05)   &   3.68(4.35)  &   3.55(4.58) \\
  $12 + \log (\rm O/H)$      &   8.65(0.13)   &    8.63(0.13)   &   8.64(0.18)  &   8.62(0.17)   &   8.57(0.34)  &   8.55(0.36) \\
  O$^{++}$/O                 &   0.80	      &    0.89	        &   0.68        &   0.66         &   0.80	 &   0.90 		 \\
  N$^{+}$/H$^+$ ($10^6$)     & 123.52(35.18)  &  148.44(40.78)  & 209.62(94.64)  & 217.78(95.49) &  92.22(89.68) &  88.29(85.88) \\
  N$^+$/O$^+$ ($10^2$)       & 135.70(21.77)  &  307.26(97.12)  & 150.04(21.69)  & 154.11(22.14) & 127.45(43.47) & 260.81(130.54) \\
  $\log {\rm N/O}$           &   0.13(0.06)   &    0.49(0.12)   &   0.18(0.06)   &   0.19(0.06)  &   0.11(0.13)  &   0.42(0.18) \\
  N/H ($10^6$)               & 604.20(225.67) & 1321.70(627.57) & 656.42(350.27) & 639.16(329.01) & 469.15(576.92) & 925.69(1280.33) \\
  $12 + \log (\rm N/H)$      &   8.78(0.14)   &    9.12(0.17)   &    8.82(0.19)  &   8.81(0.18)  &   8.67(0.35)  &   8.97(0.38) \\
  S$^+$/H$^+$ ($10^7$)       &   6.74(1.92)   &    6.54(1.82)   &    8.62(4.14)  &   8.48(4.03)  &   5.93(5.70)  &   5.38(5.18) \\
  Ne$^{++}$/H$^+$ ($10^5$)   &  21.08(11.90)  &   45.73(23.93)  &   --           &   --          &  15.57(28.52) &  20.16(36.84) \\
  Ne$^{++}$/O$^+$            &   0.59(0.15)   &    1.20(0.23)   &   --           &   --          &   0.53(0.27)  &   0.63(0.31) \\
  $\log (\rm Ne/O)$          &  -0.23(0.10)   &    0.08(0.08)   &   --           &   --          &  -0.28(0.18)  &  -0.20(0.17) \\
  Ne/H ($10^4$)              &   2.65(1.12)   &    5.15(2.07)   &   --           &   --          &   1.94(2.49)  &   2.23(3.08) \\
  $12 + \log (\rm Ne/H)$     &   8.42(0.15)   &    8.71(0.15)   &   --           &   --          &   8.29(0.36)  &   8.35(0.38) \\
  Ar$^{++}$/H$^+$ ($10^7$)   &  26.32(7.35)   &   28.24(7.66)   &  21.65(9.80)   &  20.42(9.01)  &  23.24(22.19) &  22.96(21.92) \\
  Ar/H ($10^7$)              &  49.92(16.73)  &   73.48(23.92)  &  33.65(18.28)  &  31.07(16.45) &  44.94(51.49) &  64.96(74.43) \\
  $ICF(\rm Ar)$              &   1.90	      &    2.60        &   1.55        &   1.52          &   1.93	 &   2.83	    \\
  $12 + \log (\rm Ar/H)$     &   6.70(0.13    &    6.87(0.12)  &   6.53(0.19)  &   6.49(0.18)    &   6.65(0.33) &   6.81(0.33) \\
  $\log (\rm Ar/O)$          &  -1.95(0.18)   &   -1.77(0.18)  &  -2.11(0.26)  &  -2.13(0.25)    &  -1.91(0.47) &  -1.74(0.49) \\
\hline
\end{tabular}
\end{table*}

\begin{table*}
\caption{Ionic and total abundances for the background nebula of NGC\,3603.}\label{tab:n3603}
\centering
\begin{tabular}{lrrrrrrr}
\hline
\hline\\[-11pt]
Property & NA$_{12}$ & NA$_{123}$ & NB$_{12}$ & NB$_{123}$ & NC$_{12}$ & NC$_{123}$ & NX \\
\hline\\[-11pt]
  $T$(\ion{O}{iii})\,(K) (m'd)          & 8300(400)      & 8300(700)       &      8500(400)  &      8200(500)  &       9100(400)  &     8900(500)  & --             \\
  $T$(\ion{N}{ii}) \,(K) (m'd)          & --             & --              &     11500(1600) &     10100(1200) &      10400(4000) &    10500(4900) & --             \\[10pt]
  $T$(\ion{O}{iii})\,(K)          &    8300(400)   &      8300(700)  &      8500(400)  &      8200(500)  &       9100(400)  &     8900(500)  &     8600(2000) \\
  $T$(\ion{O}{ii}) \,(K)          &    8800(500)   &      8800(700)  &     11500(1600) &     10100(1200) &      10400(4000) &    10500(4900) &     9000(2100) \\
  $T$(\ion{S}{iii})\,(K)          &    8600(500)   &      8600(700)  &      8700(500)  &      8500(500)  &       9200(500)  &     9100(500)  &     8800(2100) \\
  He$^{++}$/H$^{+}$ ($10^3$) &   1.34(1.34)   &     1.43(1.43)  &    1.10(1.10)   &    1.08(1.08)   &     0.85(0.85)   &   0.88(0.88)   &   0.00(0.00) \\
  O$^+$/H$^{+}$ ($10^5$)             &   2.78(0.75)   &     2.60(1.02)  &    1.78(0.94)   &    3.05(1.52)   &     3.17(4.98)   &   3.02(5.76)   &   9.49(10.40) \\
  O$^{++}$/H$^{+}$ ($10^5$)          &  36.99(6.65)   &    36.71(11.52) &   34.89(6.82)   &   40.89(9.25)   &    25.64(4.47)   &  27.73(5.29)   &  26.32(23.02) \\
  O/H ($10^4$)               &   3.98(0.67)   &     3.93(1.16)  &    3.67(0.69)   &    4.39(0.94)   &     2.88(0.67)   &   3.07(0.78)   &   3.58(2.53) \\
  $12 + \log (\rm O/H)$      &   8.60(0.07)   &     8.59(0.11)  &    8.56(0.07)   &    8.64(0.08)   &     8.46(0.09)   &   8.49(0.10)   &   8.55(0.23) \\
  O$^{++}$/O                 &   0.93 	      &     0.93       	&    0.95	  &    0.93	    &     0.89	       &   0.90	        &   0.74	      \\
  N$^{+}$/H$^+$ ($10^6$)     &   2.01(0.33)   &     1.80(0.42)  &    2.21(0.71)   &    4.24(1.28)   &     2.46(2.34)   &   1.94(2.24)   &   6.45(4.24) \\
  N$^+$/O$^+$ ($10^2$)       &   7.24(0.40)   &     6.92(0.52)  &   12.38(1.09)   &   13.89(1.18)   &     7.77(1.97)   &   6.41(1.97)   &   6.79(1.31) \\
  $\log {\rm N/O}$           &  -1.14(0.02)   &    -1.16(0.03)  &   -0.91(0.04)   &   -0.86(0.04)   &    -1.11(0.10)   &  -1.19(0.12)   &  -1.17(0.08) \\
  N/H ($10^6$)               &  28.78(5.10)   &    27.22(8.26)  &   45.38(9.42)   &   61.05(14.02)  &    22.38(7.70)   &  19.72(7.86)   &  24.33(17.79) \\
  $12 + \log (\rm N/H)$      &   7.46(0.07)   &     7.43(0.12)  &    7.66(0.08)   &    7.79(0.09)   &     7.35(0.13)   &   7.29(0.15)   &   7.39(0.24) \\
  S$^+$/H$^+$ ($10^7$)       &   0.81(0.15)   &     0.65(0.16)  &    0.87(0.28)   &    1.16(0.35)   &     0.95(0.88)   &   0.77(0.87)   &   2.89(1.88) \\
  S$^{++}$/H$^+$ ($10^7$)    &   59.74(17.75) &    56.78(23.36)	&   76.53(21.75)  &   88.22(26.47)  &    52.67(13.60)  &  48.52(12.87)  &  70.18(78.97) \\
  $ICF(\rm S)$               &    3.21(0.22)  &     3.35(0.23)  &    4.30(0.33)   &    3.23(0.22)   &     2.28(0.13)   &   2.48(0.15)   &   1.41(0.04) \\
  S/H ($10^7$)               &  194.28(22.22) &   192.13(26.96) &  333.10(33.57)  &  288.25(33.11)  &   122.42(15.27)  & 122.03(14.81)  & 102.67(79.05) \\
  $12 + \log (\rm S/H)$      &    7.29(0.05)  &     7.28(0.06)  &    7.52(0.04)   &    7.46(0.05)   &     7.09(0.05)   &   7.09(0.05)   &   7.01(0.25) \\
  S/O			     &    4.89(0.99)  &     4.89(1.59)  &    9.08(1.94)   &    6.56(1.59)   &     4.25(1.12)   &   3.97(1.12)   &   2.87(2.99) \\
  $\log (\rm S/O)$	     &   -1.31(0.08)  &    -1.31(0.12)  &   -1.04(0.08)   &   -1.18(0.09)   &    -1.37(0.10)   &  -1.40(0.11)   &  -1.54(0.31) \\
  Ne$^{++}$/H$^+$ ($10^5$)   &   13.65(3.12)  &    13.84(5.51)  &   13.30(3.31)   &   15.82(4.56)   &     8.96(1.99)   &   9.77(2.38)   &   8.10(9.03) \\
  Ne$^{++}$/O$^+$            &    0.37(0.02)  &     0.38(0.03)  &    0.38(0.02)   &    0.39(0.03)   &     0.35(0.02)   &   0.35(0.02)   &   0.31(0.08) \\
  $\log (\rm Ne/O)$          &   -0.43(0.02)  &    -0.42(0.04) 	&   -0.42(0.03)   &   -0.41(0.03)   &    -0.46(0.02)   &  -0.45(0.03)   &  -0.51(0.10) \\
  Ne/H ($10^4$)              &    1.47(0.26)  &     1.48(0.45)  &    1.40(0.28)   &    1.70(0.38)   &     1.01(0.24)   &   1.08(0.28)   &   1.10(0.83) \\
  $12 + \log (\rm Ne/H)$     &    8.17(0.07)  &     8.17(0.12)  &    8.15(0.08)   &    8.23(0.09)   &     8.00(0.09)   &   8.03(0.10)   &   8.04(0.24) \\
  Ar$^{++}$/H$^+$ ($10^7$)   &   23.58(3.77)  &    23.56(5.27)  &   23.89(3.71)   &   26.45(4.33)   &    19.73(2.77)   &  20.46(2.94)   &  20.74(12.96) \\
  Ar$^{+++}$/H$^+$ ($10^7$)  &    3.75(2.05)  &     3.94(2.58)  &    1.99(2.53)   &    2.29(2.90)   &     1.44(1.64)   &   1.54(1.70)   &   --       \\
  Ar/H ($10^7$)              &   27.59(4.33)  &    27.75(5.92)  &   26.09(4.53)   &   29.00(5.26)   &    21.47(3.27)   &  22.26(3.44)   &  34.68(26.01) \\
  $ICF(\rm Ar)$              &    1.01	      &     1.01	&    1.01	  &    1.01         &     1.01         &   1.01	        &   1.67	      \\
  $12 + \log (\rm Ar/H)$     &    6.44(0.06)  &     6.44(0.08)  &    6.42(0.07)   &    6.46(0.07)   &     6.33(0.06)   &   6.35(0.06)   &   6.54(0.24) \\
  $\log (\rm Ar/O)$          &   -2.16(0.09)  &    -2.15(0.14)  &   -2.15(0.10)   &   -2.18(0.11)   &    -2.13(0.11)   &  -2.14(0.12)   &  -2.01(0.34) \\
\hline
\end{tabular}
\end{table*}
\end{document}